\documentclass[]{aa}  
\usepackage{natbib} 
\bibpunct{(}{)}{;}{a}{}{,} 
\usepackage{graphicx}
\usepackage{txfonts}
\def\ax{IGR\,J18410-0535}
\def\inte{{\em INTEGRAL}}
\def\xmm{{\em XMM-Newton}}

\def\chan{{\em Chandra}}
\def\asca{{\em ASCA}}

\def\swift{{\em Swift}}

\def \inte {{$INTEGRAL$}}
\def \xmm {{\em XMM--Newton}}
\def \chandra {{$Chandra$}}

\def \hcm {\hbox {\ifmmode $ atom cm$^{-2}\else atom cm$^{-2}$\fi}}
\def \arcmin {\hbox{$^\prime$}}

\def \apjl {ApJL}
\def \apjs {ApJS}\def \aap {A\&A}

\begin{document}
   \title{\xmm\ observations of \ax:\ The ingestion of a clump by  
   a supergiant fast X-ray transient}

\author{E. Bozzo 
   \inst{1}
	\and 
   A. Giunta 
 	\inst{2} 
 	\and  
   G. Cusumano 
   \inst{3}  
	\and	
	C. Ferrigno
	\inst{1}
	\and
   R. Walter
  	\inst{1} 
  	\and 
  	S. Campana
  	\inst{4}
  	\and
  	M. Falanga
  	\inst{5}
  	\and
  	G. Israel
  	\inst{2}
  	\and
	L. Stella
	\inst{2}
}

\authorrunning{E. Bozzo et al.}
  \titlerunning{A bright flare from \ax\ }
  \offprints{E. Bozzo}

\institute{ISDC Data Center for Astrophysics, University of Geneva
	 Chemin d'\'Ecogia 16, 1290 Versoix, Switzerland.\\
	\email{enrico.bozzo@unige.ch}
         \and
          INAF - Osservatorio Astronomico di Roma, Via Frascati 33, 00044 Rome, Italy. 
          \and
          INAF - Istituto di Astrofisica Spaziale e Fisica Cosmica di Palermo, Via U. La Malfa 153, I-90146 Palermo, Italy. 
          \and 
          INAF - Osservatorio Astronomico di Brera, via Emilio Bianchi 46, I-23807 Merate (LC), Italy. 
          \and 
          International Space Science Institute (ISSI), Hallerstrasse 6, CH-3012 Bern, Switzerland.\\         
         }

 \abstract{ \ax\ is a supergiant fast X-ray transients. This subclass of supergiant X-ray binaries 
  typically undergoes few-hour-long outbursts reaching luminosities of $10^{36}$-10$^{37}$~erg/s, the occurrence 
  of which has been ascribed to the combined effect of the intense magnetic field and rotation of the compact 
  object hosted in them and/or the presence of dense structures (``clumps'') in the wind of their supergiant companion.}  
 { \ax\ was observed for 45~ks by \xmm\ as part of a program designed to study the quiescent emission of 
 supergiant fast X-ray transients and clarify the origin of their peculiar X-ray variability.}
 {We carried out an in-depth spectral and timing analysis of these \xmm\ data.} 
 { \ax\ underwent a bright X-ray flare that started about 5~ks after the beginning of the 
 observation and lasted for $\sim$15~ks. Thanks to the capabilities of the instruments on-board \xmm,\ the whole event could 
 be followed in great detail. The results of our analysis provide strong convincing evidence that the flare was 
 produced by the accretion of matter from a massive clump onto the compact object hosted in this system.} 
 {By assuming that the clump is spherical and moves at the same 
 velocity as the homogeneous stellar wind, we estimate a mass and radius   
 of $M_{\rm cl}$$\simeq$1.4$\times$10$^{22}$~g and $R_{\rm cl}$$\simeq$8$\times$10$^{11}$~cm. 
 These are in qualitative agreement with values expected from theoretical calculations.  
 We found no evidence of pulsations at $\sim$4.7~s after investigating coherent modulations 
 in the range 3.5~ms-100~s. A reanalysis of the archival \asca\ and \swift\ data of \ax,\ for which these pulsations 
 were previously detected, revealed that they were likely to be due to a statistical fluctuation and an 
 instrumental effect, respectively.}  

\keywords{X-rays: binaries - stars: individual \ax\  - stars: neutron - X-rays: stars}

\date{Received 2011 February 15; accepted 2011 May 24}

\maketitle

\section{Introduction}
\label{sec:intro}

\ax\ (=~AX\,J1841.0-0536) is a member of the supergiant fast X-ray transients (SFXT), 
a subclass of supergiant X-ray binaries (sgHMXBs) that has attracted much attention in the past 
few years because of their peculiar behavior in the X-ray domain  
\citep[see, e.g.][for an updated list of sources in this class]{walter07}. 
In contrast to previously known sgHMXBs, which are nearly persistent in 
X-rays \citep[see, e.g.][for a review]{chaty10}, SFXTs usually 
undergo few-hour long outbursts reaching luminosities of $10^{36}$-10$^{37}$~erg/s, and 
spend most of their time life in quiescence, with typical luminosities of 
10$^{32}$-10$^{33}$~erg/s \citep{sidoli10,bozzo10}. The presence of 
black-hole accretors in these sources cannot be completely ruled out. However, 
the properties of their X-ray spectra in outburst and quiescence and the detection  
of pulsations in the X-ray emission of some SFXTs, led to the conclusion that (at least)  
most of them should host neutron-star (NS) accretors 
\citep[see e.g.,][and references therein]{bozzo08}.    

\ax\ was discovered with \asca\ 
in 1994 \citep{bamba01}. During the discovery observation, the source displayed 
two bright flares with rising times shorter than 1~hr and separated by 0.6~days. 
The peak fluxes were 2.0 and 9.5$\times$10$^{-11}$~erg~cm$^{-2}$~s$^{-1}$ (2-10 keV), respectively. 
A time-resolved spectral analysis of the event revealed that the source X-ray spectrum could  
be accurately modelled with an absorbed power-law of photon index $\Gamma$=1-2 and an 
absorption column density of $N_{\rm H}$=(3.2-7.2)$\times$10$^{22}$~cm$^{-2}$. 
An iron line with a centroid energy of 
6.4~keV and an equivalent width (EW) of $\sim$0.2~keV was also required 
by the data. During the brightest of the two flares, pulsations were 
reported at a period of $\sim$4.7~s in only the 1.9-4.9~keV 
energy band. 
The lowest X-ray flux of the source measured with 
\asca\ was 2.0$\times$10$^{-12}$~erg~cm$^{-2}$~s$^{-1}$ (2-10~keV). 

\ax\ was observed with \chan\ for 20~ks on 2004 May 12 \citep{halpern04}. 
This observation captured the source with an average flux of 
4.2$\times$10$^{-12}$~erg~cm$^{-2}$~s$^{-1}$ 
(0.5-10 keV), and the corresponding X-ray spectrum  
could be described well by an absorbed power-law model 
with $\Gamma$=1.35$\pm$0.30 and 
$N_{\rm H}$=(6.1$\pm$1.0)$\times$10$^{22}$~cm$^{-2}$. 
This observation also provided an improved position of the source at 
$\alpha_{\rm J2000}$=18$^{\rm h}$41$^{\rm m}$0$\fs$54 and 
$\delta_{\rm J2000}$=-05${\degr}$35$\arcmin$46$\farcs$8 
(nominal \chan\ position accuracy  0.6$\farcs$), and 
permitted the identification of the optical and infrared counterparts of the source 
as a B1 Ib supergiant star at an estimated distance of 
3.2$^{+2.0}_{-1.5}$~kpc \citep{nespoli08}.  

The source was also captured undergoing several few-hour-long 
outbursts with MAXI (Negoro et al. 2010) and 
\inte\ \citep[1-7~hr;][]{rodriguez04,sguera06,walter07}.  
The highest flux measured by \inte\ during these events was 120~mCrab 
\citep[20-80 keV;][]{sguera06}, corresponding to roughly 
1.8$\times$10$^{-9}$~erg~cm$^{-2}$~s$^{-1}$. The source 
high energy spectrum (20-80 keV) while in outburst could be described using a hot 
blackbody model (BB) with $kT$=8-9~keV. Outside the outbursts,   
the upper limit (1$\sigma$ c.l.) to the average hard 
X-ray flux of the source was estimated to be 1~mCrab \citep[18-60 keV, corresponding 
to roughly 1.2$\times$10$^{-11}$~erg~cm$^{-2}$~s$^{-1}$,][]{filippova04}.  

\ax\ was detected a few times in outburst by \swift\,/BAT, and only 
once \swift\ performed a slew on the source to point it with 
the narrow field instrument XRT \citep{depasquale10,romano10,romano10b}. 
On that occasion, the \swift\,XRT spectrum 
was closely described by an absorbed power-law model with 
$\Gamma$=0.7$^{+0.5}_{-0.4}$ and $N_{\rm H}$=3$^{+2}_{-1}$$\times$10$^{22}$~cm$^{-2}$. 
The average 2-10 keV unabsorbed flux was 7$\times$10$^{-10}$~erg~cm$^{-2}$~s$^{-1}$. 
No evidence for pulsations was found. 
\ax\ was not detected in a high X-ray activity phase 
during the one-year monitoring (total net on-source exposure time 96.5~ks) 
with \swift\,/XRT carried out by \citet{romano09}. In the data acquired during 
this monitoring, the authors identified four different states 
for the X-ray emission of \ax:\ the ``high'', ``medium'', ``low'', and ``very low'' states. 
In the high state, the source X-ray flux was 8.0$\times$10$^{-11}$~erg~cm$^{-2}$~s$^{-1}$ (2-10 keV) and 
the corresponding spectrum is closely described by an absorbed power-law model 
with $\Gamma$=1.1$\pm$0.1 and $N_{\rm H}$=(2.5$\pm$0.3)$\times$10$^{22}$~cm$^{-2}$. 
The medium state was characterized by a flux of 3.4$\times$10$^{-11}$~erg~cm$^{-2}$~s$^{-1}$, 
a power-law photon index of 1.3$\pm$0.2, and $N_{\rm H}$=(3.5$\pm$0.5)$\times$10$^{22}$~cm$^{-2}$. 
For the low and very low states, the authors estimated fluxes of 
1.1$\times$10$^{-11}$~erg~cm$^{-2}$~s$^{-1}$ and 6.0$\times$10$^{-13}$~erg~cm$^{-2}$~s$^{-1}$, respectively. 
The corresponding power-law photon indices and absorption column densities were 
$\Gamma$=1.5$\pm$0.1, 0.6$\pm$0.4 and $N_{\rm H}$=(3.5$\pm$0.5)$\times$10$^{22}$~cm$^{-2}$, 
(0.6$\pm$0.4)$\times$10$^{22}$~cm$^{-2}$. 
\citet{sidoli08} also reported the detection of pulsations at $\sim$4.7~s 
from \ax\ in some of the \swift\,/XRT data.  
A possible association between \ax\ and the transient MeV EGRET source 3EG\,J1837-0423 
was suggested by \citet{sguera09}. 

In this paper, we report on a 45~ks long observation of the source with \xmm.\ 
This observation was obtained as part of our program aimed at studying  
the quiescent emission of SFXTs \citep[see also][]{bozzo10,bozzo09,bozzo08}. 
During the \xmm\ observation, \ax\ was caught undergoing a bright X-ray flare, 
lasting roughly 15~ks. The source reached 
a maximum flux of 3.4$\times$10$^{-10}$erg~cm$^{-2}$~s$^{-1}$ (1-10~keV) at the peak of the flare and then decayed 
to a very low quiescent level (8.8$\times$10$^{-14}$erg~cm$^{-2}$~s$^{-1}$). The total dynamic range 
in the X-ray flux was thus $\gtrsim$4$\times$10$^{3}$. In Sect.~\ref{sec:xmmdata}, we give 
the details of our data analysis and provide all the results in 
Sect.~\ref{sec:xmmresultsspectra} and \ref{sec:xmmresultstiming}. 
In the \xmm\ data,  
no pulsation at 4.7~s could be detected, and the derived upper limits were significantly tighter  
than the pulsation amplitude reported previously with \asca\ and \swift.\  
Motivated by these results, we also reanalyzed all the \asca\ and \swift\ data of the source 
and show that the previously reported 4.7~s pulsations were probably due a statistical fluctuation (\asca\ ) and 
an instrumental effect (\swift\ ), respectively (Sects.~\ref{sec:ascadata} and \ref{sec:swiftdata}).  
Our discussion and conclusions are reported in Sects.~\ref{sec:discussion} and \ref{sec:conclusion}.

\section{ \xmm\ data analysis}
\label{sec:xmmdata}

\xmm\ \citep{jansen01} observed \ax\ from 55270.5484 MJD to 55271.0923 MJD (exposure time 
$\sim$45~ks).  
The Epic-pn camera \citep{struder01} was operated in full frame, 
while the Epic-MOS1 and Epic-MOS2 cameras \citep{turner01} operated in small window and 
timing mode, respectively. This particular set-up of the instruments was chosen 
in order to have the capabilities required to study the source over a wide range  
of X-ray flux. 
We processed the \xmm\ observation data files 
with the two pipelines {\sc epproc} and {\sc emproc} in order to produce 
Epic-pn and Epic-MOS cleaned event files, respectively (SAS v.10.0.1). 
\begin{figure}
\centering
\includegraphics[scale=0.35,angle=-90]{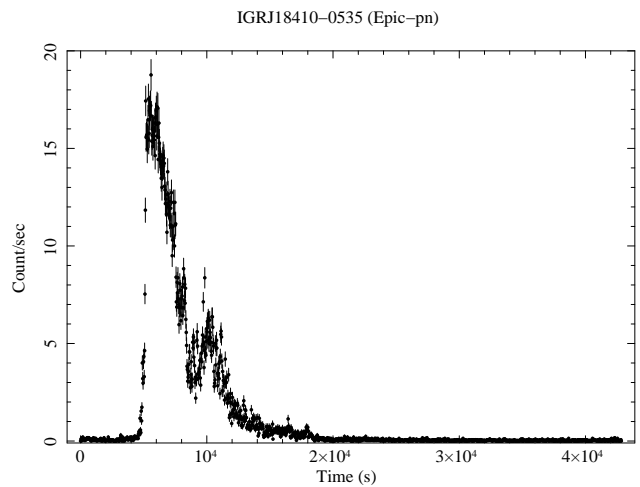}
\caption{ \xmm\ Epic-pn lightcurve of the observation of \ax\ 
(0.3-12~keV, not corrected for background and pile-up). The start time 
is MJD 55270.5896 (bin time 30~s).} 
\label{fig:total_lcurve} 
\end{figure}
No time intervals were found 
to be affected by a high background and we thus retained for the subsequent  
analysis the entire exposure time available for the three instruments. 
We used the energy range 0.15-15~keV for the Epic-pn and 0.3-12 keV for the two 
Epic-MOS cameras. 
\begin{figure}
\centering
\includegraphics[scale=0.35,angle=-90]{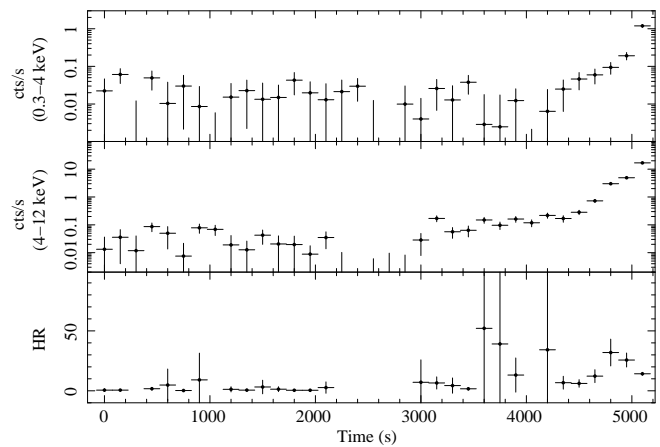}
\caption{Rise of the flare observed during the \xmm\ observation 
of \ax\ with the Epic-pn camera. The start time of the lightcurve is the 
same as that in Fig.~\ref{fig:total_lcurve}, 
and the binning time is 150~s. The upper panel shows the source lightcurve 
in the soft energy band (0.3-4~keV), the middle panel in the hard energy band 
(4-12~keV), and the bottom panel the hardness ratio (HR, defined as the ratio of  
the source count-rate in the soft to hard energy band versus time.) }
\label{fig:rise} 
\end{figure} 
\begin{figure}
\centering
\includegraphics[scale=0.35,angle=-90]{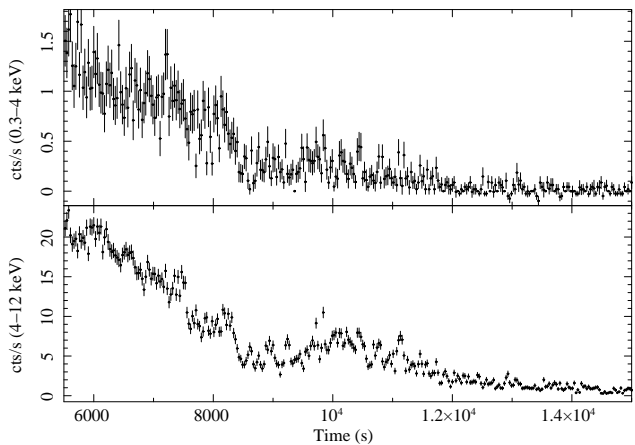}
\caption{Details of the Epic-pn lightcurve during the drop in count rate that occurred 
around $t$=8500-9500~s in Fig.~\ref{fig:total_lcurve} ($t$=0 would correspond to the the start 
time of the lightcurve in Fig.~\ref{fig:total_lcurve}; time bin is 30~s). The upper (lower) panel 
shows the source lightcurve in the 0.3-4~keV (4-12~keV) energy band.} 
\label{fig:hole} 
\end{figure}
\begin{figure}
\centering
\includegraphics[scale=0.35,angle=-90]{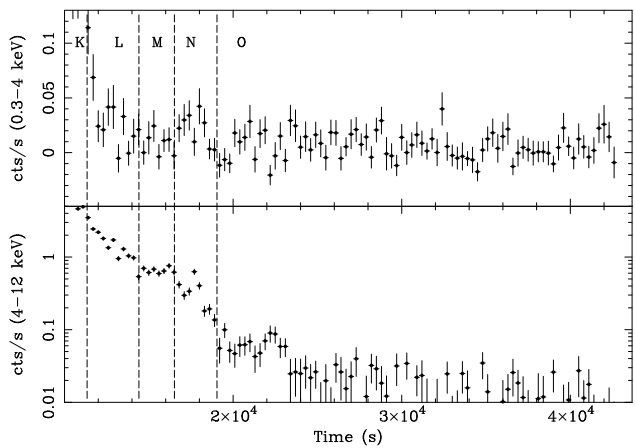}
\caption{Final part of the decay after the flare 
observed by \xmm.\ The upper (lower) panel shows the source lightcurve 
in the 0.3-4~keV (4-12~keV) energy band. We marked with vertical dashed lines 
the time intervals during which major changes in the source spectrum occurred 
(spectra K,L,M,N,O in Table~\ref{tab:fit} and Fig.~\ref{fig:decayspectra}).}  
\label{fig:decay} 
\end{figure} 
\ax\ displayed a large dynamical range in the X-ray flux during the \xmm\ 
observation (see Fig.~\ref{fig:total_lcurve}); therefore the  
source and background extraction regions were 
chosen in the different time intervals so as to maximize the signal-to-noise ration 
(S/N) of the data (see next section). 
During the periods of greatest X-ray emission from the source, 
the Epic-pn data suffered significant 
pile-up. We accounted for this problem by using in these cases annular 
extraction regions for the 
source in which the appropriate innermost portion of the instrument point 
spread function (PSF) was removed\footnote{See also  
http://xmm.esac.esa.int/external/xmm\_user\_support 
/documentation/uhb/index.html.}. We checked that the removal of  
pile-up was effective by using the {\sc SAS} tool  
{\sc epatplot} and comparing the results from the Epic-pn camera 
with those obtained with the Epic-MOS1 and Epic-MOS2. The effect of pile-up 
in these cases was much reduced since the cameras were operated in small window and timing 
mode, respectively. 
Where required, we corrected Epic images to remove the 
out-of-time (OoT) events\footnote{See http://xmm.esa.int/sas/current/documentation/threads/
EPIC\_OoT.shtml.} and checked that this problem 
did not significantly affect the spectra. 
All the Epic lightcurves were barycentered by using the SAS tool {\sc barycorr}, and then 
corrected for instrumental vignetting, dead time, and 
PSF losses by using the tool {\sc epiclccorr}. 
Given the lower count rate and S/N of the Epic-MOS1 and Epic-MOS2 cameras  
compared to those obtained from the Epic-pn (especially outside the flare), the former instruments 
did not contribute significantly to the spectral analysis. Therefore, we report in the following 
sections and in Fig.~\ref{fig:total_lcurve} only the results obtained from the Epic-pn camera. 

\subsection{Lightcurve analysis}
\label{sec:lightcurvespectra}

During the \xmm\ observation, \ax\ underwent 
a bright X-ray flare that started about 5~ks after the beginning 
of the observation and lasted for $\sim$15~ks. 
The lightcurve of the entire \xmm\  
observation in the 0.3-12~keV energy band 
is shown in Fig.~\ref{fig:total_lcurve}. 
\begin{figure}
\centering
\includegraphics[scale=0.35,angle=-90]{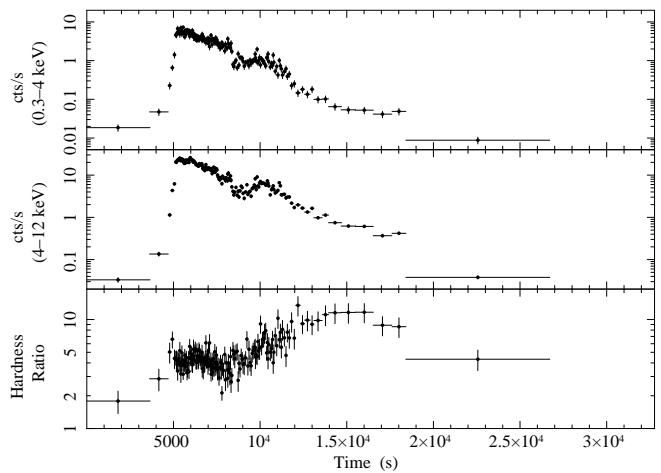}
\caption{The same as Fig.~\ref{fig:total_lcurve}, but here we used an adaptive 
rebinning with S/N=5. The upper panel shows the source lightcurve in the 0.3-4~keV energy band, 
the middle panel the source lightcurve in the 4-12~keV energy band, and the lower panel 
the hardness ratio versus time (defined as in Fig.~\ref{fig:rise}).}  
\label{fig:hr} 
\end{figure}

The lowest source count-rate was recorded during the latest 
20~ks of observation at a level of (4.0$\pm$1.9)$\times$10$^{-3}$~cts/s 
(estimated from a 5000~s binned lightcurve that had background subtracted; 
all uncertainties in this paper are given at 90\% c.l., unless 
otherwise indicated). During the peak of the flare, the corresponding 
highest count-rate was 31.2$\pm$1.0~cts/s 
(determined from a 30~s binned lightcurve, which was  
background subtracted and corrected for pile-up). 
In Fig.~\ref{fig:rise}, we show an enlargement of the lightcurve during the first 5~ks of the \xmm\ observation. 
Here, a sudden rise of the source X-ray emission was recorded by the 
Epic-pn camera $\sim$3~ks after the beginning of the observation. In the following $\sim$2~ks,  
the source reached the highest count-rate, and then began to decrease 
at $t$$\sim$6000~s.  
Interestingly a drop in the source count rate was recorded around $t$=8500~s in both 
the soft and hard energy band. A zoom in of this part of the lightcurve is provided in 
Fig.~\ref{fig:hole}. A detailed lightcurve of the final part of the flare decay 
is shown in Fig.~\ref{fig:decay}. 
\begin{figure}
\centering
\includegraphics[scale=0.35,angle=-90]{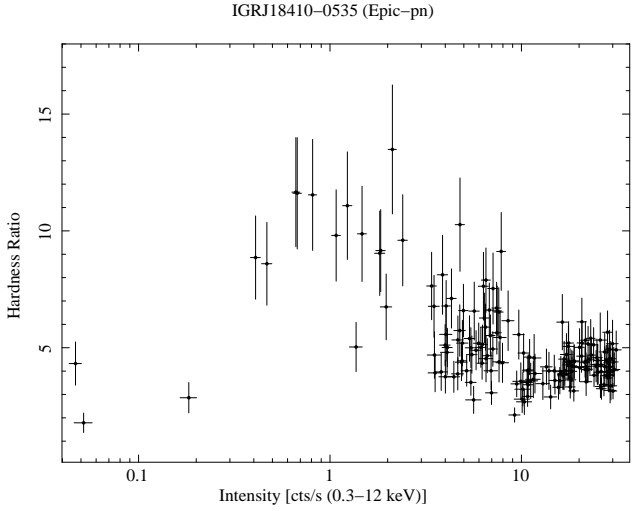}
\caption{ Hardness-intensity diagram of \ax\ obtained by using the \xmm\ Epic-pn 
observation. The hardness ratio of the source is defined as in Fig.~\ref{fig:rise}.} 
\label{fig:intensitydiagram} 
\end{figure}
We also show in Fig.~\ref{fig:hr} the source lightcurves 
in the 0.3-4~keV and 4-12~keV energy bands, where  
an adaptive rebinning of the data was used in 
order to achieve in each time bin S/N=5 and  
estimate the corresponding source hardness ratio (HR).  
The same lightcurves and adaptive rebinning were used 
to produce the hardness ration versus intensity diagram shown in Fig.\ref{fig:intensitydiagram}
\citep[see also][]{bozzo10}. 

From Figs.~\ref{fig:hr} and \ref{fig:intensitydiagram}, 
we can see that the HR of the source underwent dramatic changes  
in time; a non-monotonic trend appears when the HR is plotted as a function 
of the total source intensity. Given these findings, we performed in 
Sect.~\ref{sec:xmmresultsspectra} below a time-resolved spectral analysis of the data.

\subsection{Spectral analysis}
\label{sec:xmmresultsspectra}

\begin{figure}
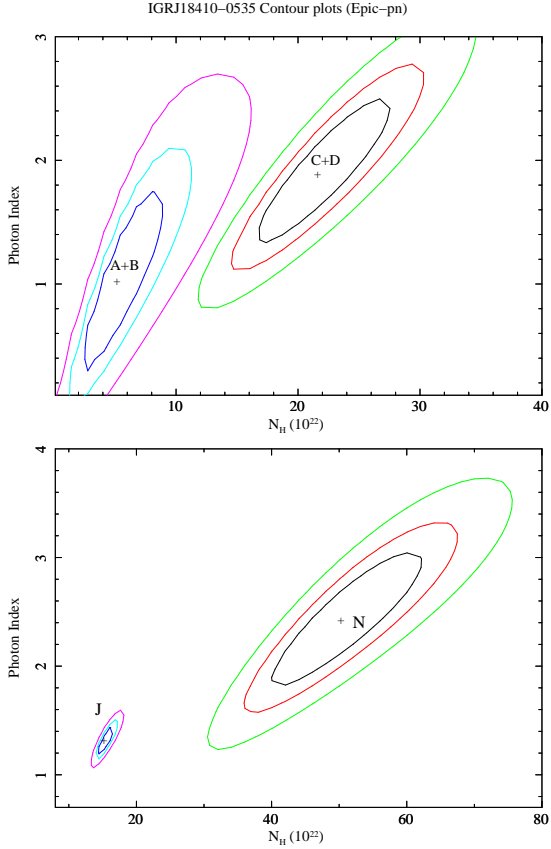

\centering 
\includegraphics[scale=0.3,angle=-90]{16726fig9.ps}
\includegraphics[scale=0.3,angle=-90]{16726fig10.ps}
\caption{{\it Upper panel}: Contour plots for the spectral parameters 
measured from the time intervals A+B and C+D (see Table~\ref{tab:fit}). 
The contours correspond to the 68\%, 90\%, and 99\% c.l.
{\it Lower panel}: Same as above but for the J and 
N spectra in Table~\ref{tab:fit}.}
\label{fig:contours} 
\end{figure}

To search for spectral changes, we first 
accumulated X-ray spectra on time intervals of $\sim$few hundreds to $\sim$few thousands 
seconds (depending on the source count-rate), following the trend of the HR  
(see Fig.~\ref{fig:hr}). Nearby time intervals were then combined in order to extract a lower number 
of spectra that could maximize the evidence for changes in the properties of the source X-ray emission. 
Our optimal separation of the time intervals is reported in Table~\ref{tab:fit}.   
All these spectra were rebinned to have at least 15, 20, or 25 
photons per bin (depending on the source intensity) and prevent an oversampling of the
energy resolution of the instruments by more than a factor of three. Spectra with lower quality statistics were 
rebinned to have at least 5 photons per bin and were fit using C-statistics 
\citep{cash79}. For the spectral fits, we used an absorbed power-law model\footnote{We used the model phabs*pow in 
{\sc xspec}.}. None of the spectra could be convincingly better fit with a blackbody (BB) or a cut-off power-law model 
({\sc cutoffpl} in {\sc Xspec}). In particular, the latter provided in most cases peculiar values  
of the model parameters (negative power-law photon indices and values of the high energy cut-off 
largely inconsistent between the different spectra). 
\begin{figure}
\centering
\includegraphics[scale=0.52]{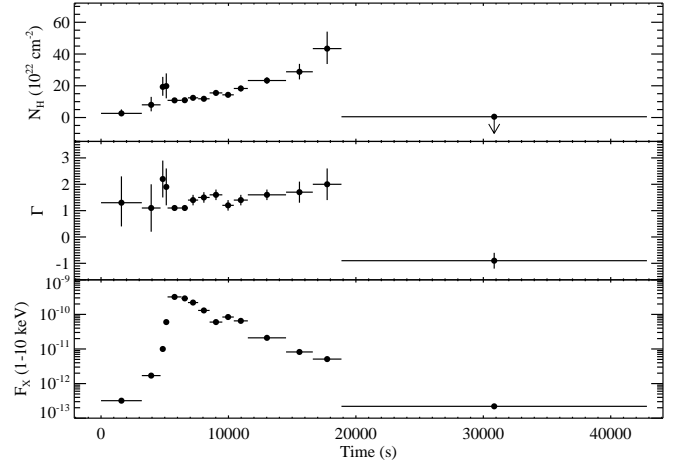}
\caption{Spectral parameters of \ax\ measured during the different time 
intervals reported in Table~\ref{tab:fit}. All the errors are at 90\% c.l. 
Arrows indicate 90\% c.l. upper limits.}
\label{fig:parameters} 
\end{figure}
A plot of the evolution of the spectral parameters of the source 
in the \xmm\ observation is provided in Fig.~\ref{fig:parameters}.  
The power-law photon index remained fairly constant up to $t$$\simeq$19000 s,  
whereas the absorption column density underwent dramatic changes (Fig.~\ref{fig:parameters}). 
\begin{figure}
\centering
\includegraphics[scale=0.52,angle=-90]{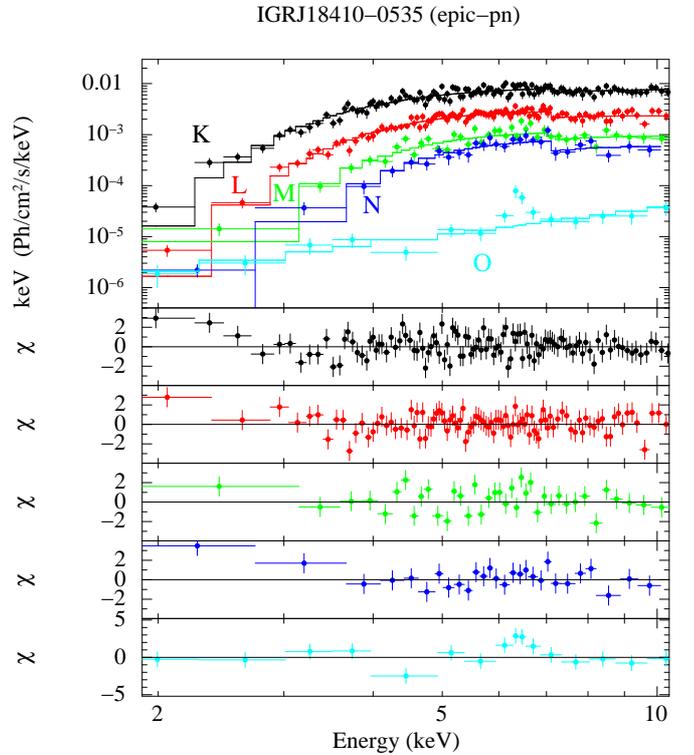}
\caption{ \xmm\ Epic-pn spectra of \ax\ extracted during the latest part of the 
flare decay (see Table~\ref{tab:fit} and Fig.~\ref{fig:decay}). 
All the spectra were fit with a simple absorbed power-law model (the residuals 
from these fits are shown for each of the five spectra).}  
\label{fig:decayspectra} 
\end{figure}
\begin{table*}[t]
\tiny
\centering
\caption{Best-fit parameters of \ax\ during different time intervals 
of the \xmm\ observation (see Fig.~\ref{fig:total_lcurve} and the note below 
the table).}
\begin{tabular}{lllllllllll}
\hline
\hline
\noalign{\smallskip} 
Interval & Tstart & Tstop & Exp. & $N_{\rm H}$  & $\Gamma$  & $E_{\rm line}$ & $EW_{\rm line}$ & $F_{\rm obs}$ & $F_{\rm unabs}$ & $\chi_{\rm red}^2$/d.o.f. \\
\noalign{\smallskip}
         & (s)    &  (s)  & (ks) & (10$^{22}$~cm$^{-2}$) &  & (keV) & (keV) & (erg~cm$^{-2}$~s$^{-1}$) & (erg~cm$^{-2}$~s$^{-1}$)  & (C-statistics/d.o.f.) \\
\hline
\noalign{\smallskip} 
A & 0 & 3200 &  2.7  & 2.6$^{+2.5}_{-1.7}$ & 1.3$^{+1.0}_{-0.9}$  & --- & --- & 3.2$\times$10$^{-13}$ & 4$\times$10$^{-13}$ & (12.4/11) \\
\noalign{\smallskip}
B & 3200 & 4670 & 1.2 & 8.0$^{+5.0}_{-4.1}$ & 1.1$\pm$0.9 &  --- & --- & 1.7$\times$10$^{-12}$  & 2.8$\times$10$^{-12}$  & (14.4/18) \\         
\noalign{\smallskip}
C & 4670 & 5020 &  0.35 & 19.3$^{+6.3}_{-5.6}$ & 2.2$\pm$0.7 &  --- & --- & 1.0$\times$10$^{-11}$ & 5.0$\times$10$^{-11}$ & (51.0/47) \\
\noalign{\smallskip}
D & 5020 & 5220 & 0.17 & 19.8$^{+8.0}_{-7.7}$ & 1.9$\pm$0.7 &  --- & --- & 6.0$\times$10$^{-11}$ & 2.4$\times$10$^{-10}$ & 0.8/12 \\
\noalign{\smallskip}
E & 5220 & 6320 & 0.96 & 10.8$^{+0.7}_{-0.6}$ & 1.1$\pm$0.1 &  6.56$\pm$0.05 & 0.06 & 3.2$\times$10$^{-10}$ & 6.0$\times$10$^{-10}$ & 1.0/149 \\         
\noalign{\smallskip} 
F & 6320 & 6820 & 0.44 & 10.9$^{+1.2}_{-1.1}$ & 1.1$\pm$0.1 &  --- & --- & 2.9$\times$10$^{-10}$ & 5.3$\times$10$^{-10}$ & 0.9/109 \\
\noalign{\smallskip}
G & 6820 & 7620 & 0.7 & 12.4$^{+1.3}_{-1.2}$ & 1.4$\pm$0.2 &  --- & --- & 2.2$\times$10$^{-10}$ & 4.4$\times$10$^{-10}$ & 1.1/108 \\
\noalign{\smallskip}
H & 7620 & 8520 & 0.8 & 11.8$^{+1.5}_{-1.3}$ & 1.5$\pm$0.2 &  --- & --- & 1.3$\times$10$^{-10}$ & 2.9$\times$10$^{-10}$ & 1.1/73 \\         
\noalign{\smallskip}
I & 8520 & 9520 & 0.9 & 15.5$^{+1.6}_{-1.5}$ & 1.6$\pm$0.2 &  6.32$\pm$0.05 & 0.10 & 6.2$\times$10$^{-11}$ & 1.8$\times$10$^{-10}$ & 0.8/107 \\
\noalign{\smallskip}
J & 9520 & 10420 & 0.8 & 14.3$^{+1.4}_{-1.5}$ & 1.2$\pm$0.2 &  --- & --- & 8.4$\times$10$^{-11}$ & 1.9$\times$10$^{-10}$ & 0.8/110 \\
\noalign{\smallskip}
K & 10420 & 11520 & 1.0 & 17.0$\pm$1.5 & 1.3$\pm$0.2 &  --- & --- & 6.5$\times$10$^{-11}$ & 1.6$\times$10$^{-10}$ & 1.1/114 \\
\noalign{\smallskip}  
L & 11520 & 14520 & 2.6 & 23.3$\pm$2.0 & 1.6$\pm$0.2 &  --- & --- & 2.3$\times$10$^{-11}$ & 8.5$\times$10$^{-11}$ & 1.0/102 \\
\noalign{\smallskip}
M & 14520 & 16620 & 1.8 & 28.0$^{+5.0}_{-4.8}$ & 1.7$\pm$0.4 &  --- & --- & 8.2$\times$10$^{-12}$ & 3.3$\times$10$^{-11}$ & 1.2/58 \\
\noalign{\smallskip}
N & 16620 & 18870 & 1.8 & 43.4$^{+10.7}_{-9.7}$ & 2.0$\pm$0.6 &  --- & --- & 5.1$\times$10$^{-12}$ &  5.3$\times$10$^{-11}$ & 1.4/36 \\
\noalign{\smallskip}
O & 18870 & 42840 &  20.8 & $<$0.5 & -0.9$\pm$0.3 & 6.36$\pm$0.06 & 1.1 & 2.2$\times$10$^{-13}$ &   2.2$\times$10$^{-13}$ & 1.0/13 \\
\noalign{\smallskip}
O$_{\rm 1}$ & 18870 & 22820 &  3.4 & $<$8.0 & -1.8$^{+0.6}_{-0.4}$ & 6.33$\pm$0.06 & 1.2 & 9.7$\times$10$^{-13}$ & 9.8$\times$10$^{-13}$  & 0.8/5 \\
\noalign{\smallskip}
O$_{\rm 2}$ & 22820 & 42840 &  17.4 & $<$1.4 & 0.3$^{+0.5}_{-0.6}$ & --- & --- & 8.8$\times$10$^{-14}$ & 9.1$\times$10$^{-14}$  & (12.5/10) \\
\noalign{\smallskip}
\hline
\hline
\noalign{\smallskip}
\end{tabular}
\tablefoot{The continuum spectral model is an absorbed power law. Here, $N_{\rm H}$ is the absorption column density, 
$\Gamma$ is the power-law photon index, $E_{\rm line}$ is the energy of the centroid of the iron line, and $EW_{\rm line}$ the 
corresponding equivalent width. $F_{\rm obs}$ ($F_{\rm unabs}$) is the absorbed (unabsorbed) flux in the 1-10~keV band. 
In the last column we report the value of the $\chi_{\rm red}^2$/d.o.f or the corresponding value of the C-statistics/d.o.f, 
for each spectrum.}
\label{tab:fit}
\end{table*}  
In particular, we measured a large increase (factor of $\sim$10) in the absorption column density 
from the beginning of the observation up to the peak of the flare 
($N_{\rm H}$$\sim$2$\times$10$^{23}$~cm$^{-2}$ at $t$$\simeq$5000~s). A rapid 
drop in the $N_{\rm H}$ then occurred to $\sim$10$^{23}$~cm$^{-2}$. At the peak of the flare, 
we also detected a significant iron line with an energy of $\sim$6.6~keV and an EW$\sim$60~eV 
(see Table~\ref{tab:fit}).  
The absorption column density remained virtually constant at $\sim$10$^{23}$~cm$^{-2}$ for about 
4~ks after the source reached the highest X-ray emission level; a rapid rise in column density then occurred  
until $t$$\simeq$19000~s. To test the significance of the measured variation 
in the absorption column density, we report in Fig.~\ref{fig:contours} the $N_{\rm H}$-$\Gamma$ 
parameters confidence contours for some relevant spectra.   
In the upper panel of this figure, we show the parameter contours obtained by summing the spectra from intervals  
C and D and A and B, which displayed very similar values of the power-law photon 
index and absorption column density before and during the rise of the flare, respectively (see Table~\ref{tab:fit}). 
The contours correspond to 68\%, 90\%, and 99\% c.l.. A very significant ($>$3~$\sigma$) 
increase in the absorption column density at the onset of the flare is clearly seen. 
The lower panel of Fig.~\ref{fig:contours} shows the case of the J and N spectra: the increase in the 
absorption column density toward the end of the flare is also highly significant.  
During the apparent drop in count rate around $t$=8500-9500~s, no particular 
change in the spectral continuum is seen, but a significant 
iron line appears with a centroid energy of $\sim$6.4~keV 
(see spectrum I in Table~\ref{tab:fit}; throughout this paper, we use 
narrow lines for the spectral fits and fixed their width to 0 in {\sc Xspec}). 

\begin{figure}
\centering 
\includegraphics[scale=0.35,angle=-90]{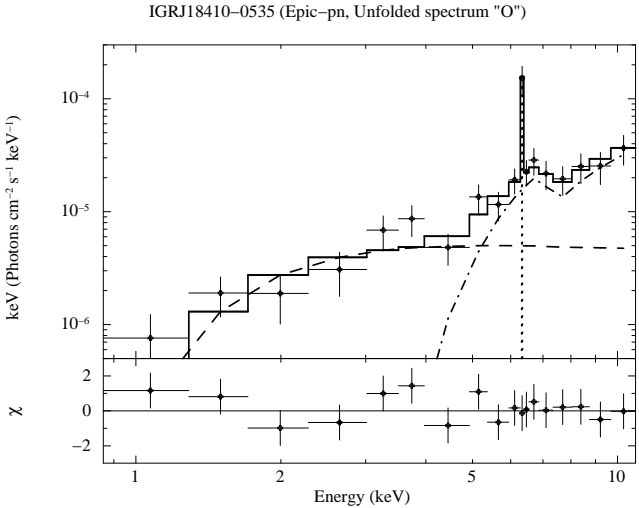}
\caption{Unfolded Epic-pn spectrum of \ax\ during the interval O in Table~\ref{tab:fit}. 
The best-fit model in {\sc Xspec} here is phabs1*(pow1+gauss+phabs2*pow2). The value of 
phabs1 was fixed in the fit to 1.8$\times$10$^{22}$~cm$^{-2}$ and the photon index 
of the two power laws was forced to be the same (see Sec.~\ref{sec:xmmresultsspectra} for details). The dashed line 
corresponds to the less absorbed power-law component in the fit, and the dot-dashed line to the more absorbed one. 
The dotted line indicates the iron line component at $\sim$6.4~keV. The bottom panel shows the residuals from the fit.}  
\label{fig:unfolded} 
\end{figure}

The properties of the X-ray emission of the 
source changed sharply around $t$$\simeq$19000~s: at this time, the source underwent a further decrease in the X-ray 
flux, and simultaneously its spectrum flattened and a prominent iron emission line appeared around 6.4~keV. We show in 
Fig.~\ref{fig:decayspectra} the detail of the evolution of the source spectrum with time during the latest 30~ks 
of the \xmm\ observation. 
A fit to the spectrum from the O interval with a simple absorbed power-law model 
gave unacceptable results ($\chi^2_{\rm red}$/d.o.f.=2.1/15). 
Adding an iron line with an energy centroid of 
$E_{\rm line}$=6.36$^{+0.06}_{-0.05}$~keV significantly improved the fit 
($\chi^2_{\rm red}$/d.o.f.=1.0/13). The normalization and equivalent width of the line 
were (2.7$\pm$1.0)$\times$10$^{-6}$ and 1.1~keV, respectively. 
From this spectrum, only an upper limit 
on $N_{\rm H}$ could be obtained (see Table~\ref{tab:fit}); this  
was significantly  lower than that expected in the direction of the source 
\citep[$\sim$1.8$\times$10$^{22}$~cm$^{-2}$,][]{dickey90}.  
Even though this model gave a statistically acceptable fit to the spectrum O, 
the very low value of the absorption column density and the unphysical negative power-law photon index suggest that 
other spectral models cannot be excluded. As we discuss in Sect.~\ref{sec:discussion}, a more physical interpretation of the 
X-ray emission of the source in this time interval is based on a model\footnote{We also performed a fit using the 
reflection model {\sc pexrav} in {\sc Xspec}, but the poor statistics  
of the O spectrum did not permit to derive satisfiable constraints on the model parameters. We thus do not discuss this model 
in further details.} comprising two power-law components affected by two different absorption column densities 
plus the iron line, i.e. {\sc phabs1*(pow1+gauss+phabs2*pow2)} in {\sc Xspec}. 
In the fit, we fixed phabs1 to the Galactic value of the absorption column density and constrained the photon indices 
of the two power-laws to be the same. The fit gave $\chi^2_{\rm red}$/d.o.f.=0.8/12, $\Gamma$=0.9$^{+1.8}_{-0.5}$ (in agreement 
with that measured during the other time intervals in Table~\ref{tab:fit}), $E_{\rm line}$=6.36$\pm$0.06~keV, 
$EW$=0.7~keV, and a value for the second absorption component of (85$_{-33}^{+48}$)$\times$10$^{22}$~cm$^{-2}$. 
In Fig.~\ref{fig:unfolded} we report the unfolded spectrum of interval O together with the different spectral component 
used in this model. 

\begin{figure*}
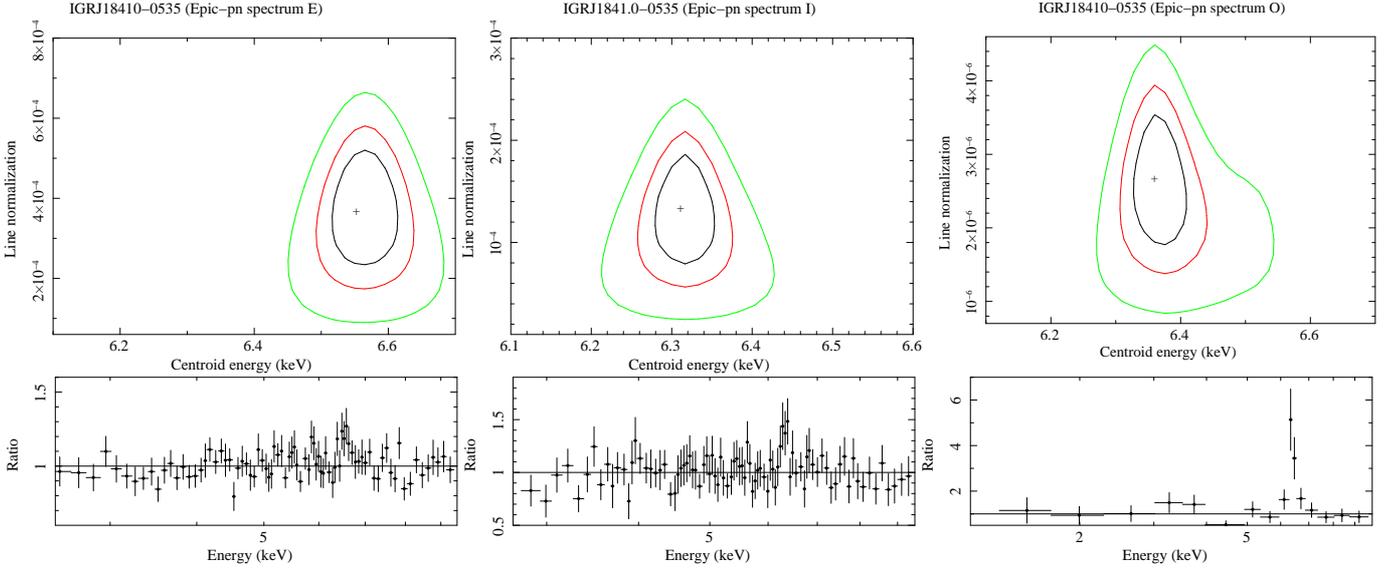

\centering
\includegraphics[scale=0.31,angle=-90]{16726fig13.ps}
\includegraphics[scale=0.31,angle=-90]{16726fig14.ps}
\includegraphics[scale=0.30,angle=-90]{16726fig15.ps} 
\includegraphics[scale=0.31,angle=-90]{16726fig16.ps}
\includegraphics[scale=0.31,angle=-90]{16726fig17.ps}
\includegraphics[scale=0.31,angle=-90]{16726fig18.ps}
\caption{{\it Top panels}: Contour plots of the iron line centroid energy vs. normalization 
measured from the spectrum E (left), I (middle), and O (right). 
The contours correspond to the 68\%, 90\%, and 99\% c.l. 
There is a marginal indication of a change in the centroid energy of the 
line between the time interval E and the other two. 
{\it Bottom panels}: The ratio of the data to the model of the spectra extracted 
during the time interval E (left), I (middle), and O (right). All these spectra 
were fit by using only an absorbed power-law model in order to highlight the presence of the 
iron line in the residuals from the fit.}   
\label{fig:lines} 
\end{figure*}

In Fig.~\ref{fig:lines}, we also show the contours plots of the iron line centroid energy versus its normalization  
for the three detections in the E, I, and O spectra. We find marginal evidence for a decrease in the 
line centroid with time, from the peak of the flare (spectrum E) to the quiescent level (spectrum O). 
No other convincing evidence of the iron line could be found in the spectra 
extracted from different time intervals or from merging them in different combinations. 
The interpretation of all the above findings is discussed in Sect.~\ref{sec:ingestion}. 

We also performed a separate spectral analysis of the source X-ray emission during 
the first $\sim$3~ks and the last $\sim$17~ks of the time interval O in order to 
investigate the cause of the drop in the source count-rate visible 
in Fig.~\ref{fig:decay} around $t$=22820. We indicate these two additional spectra 
with O$_{\rm 1}$ (from $t$=18870 to $t$=22820) and O$_{\rm 2}$ 
(from $t$=22820 to $t$=42840) in Table~\ref{tab:fit}. 

The O$_{\rm 1}$ spectrum comprised 151 counts and was dominated 
by the presence of the iron line. 
A fit with a simple absorbed power-law did 
not provide an acceptable fit ($\chi^2_{\rm red}$/d.o.f.=2.9/7), whereas adding a 
Gaussian line to the spectral model significantly improved the fit ($\chi^2_{\rm red}$/d.o.f.=0.8/5). 
We measured a power-law photon index of $\Gamma$=-1.8$^{-0.4}_{+0.6}$, and a centroid energy for the line 
of 6.33$^{+0.06}_{-0.04}$~keV. The normalization of the line was (1.1$\pm$0.4)$\times$10$^{-5}$ and the estimated 
EW 1.2~keV (compatible with that found by using the entire time interval O). Only an upper limit 
to the absorption column density of $N_{\rm H}$$<$8.0$\times$10$^{22}$~cm$^{-2}$ could be obtained. 

Given the measured negative value of the parameter $\Gamma$, we also fit the spectrum 
O$_{\rm 1}$  by using the double power-law model discussed above and checked the consistency of the different 
model parameters with those determined before for the spectrum O.  
This gave $\Gamma$=0.2$_{-1.5}^{+1.0}$ and an absorption column density for the most 
extinguished power-law component of 82$_{-42}^{+52}$$\times$10$^{22}$~cm$^{-2}$ (the absorption column 
density of the other component was fixed to the Galactic value). The centroid energy of the line 
and its EW was found to agree with those reported before for the spectrum O. 
The estimated 1-10~keV X-ray flux was 9.7$\times$10$^{-13}$~erg~cm$^{-2}$~s$^{-1}$. 

During the time interval O$_{\rm 2}$, \xmm\ recorded only 83 counts from the source. The 
corresponding spectrum could be closely fit with a simple absorbed power-law model (see Table~\ref{tab:fit}). 
The quality of the fit did not change significantly (C-statistics 14.5/11) when the power-law photon index 
was constrained to be equal to that measured for the O spectrum ($\Gamma$=1.0); the corresponding value  
of the absorption column density was $N_{\rm H}$=(1.6$_{-1.0}^{+2.0}$)$\times$10$^{22}$~cm$^{-2}$, 
compatible with the expected Galactic value in the direction of the source. Given these results, 
we did not attempt to fit this spectrum with a double power-law model.  

Adding to the spectral model an iron line with an energy fixed at $E_{\rm line}$=6.36~keV provided an upper limit 
(90\% c.l.) to its normalization and EW of 1.8$\times$10$^{-6}$ and 1.9~keV, respectively. While the upper limit 
to the EW is consistent with that estimated for the O spectrum, the normalization is slightly lower. 
However, it is still compatible with the value expected because of the decrease in the flux in the two spectra. 

\begin{figure}
\centering 
\includegraphics[scale=0.35,angle=-90]{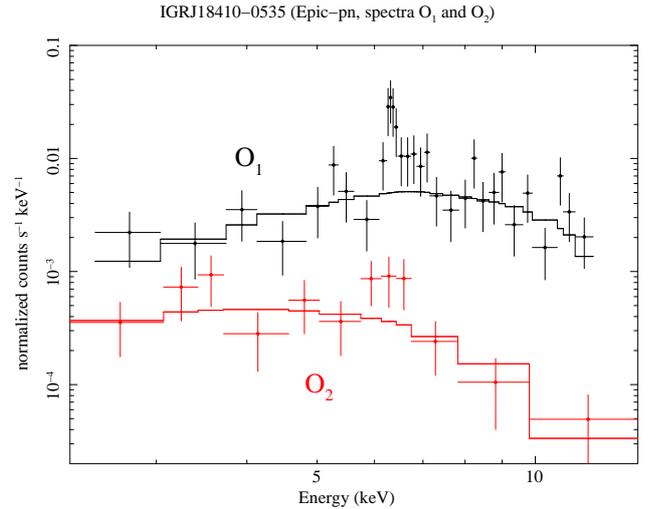}
\caption{The two spectra O$_{\rm 1}$ and O$_{\rm 2}$ extracted during the 
fist $\sim$3~ks and the last $\sim$17~ks of the time interval O in Table~\ref{tab:fit}. 
To facilitate the visual comparison, we used in this figure a grouping of 5 photons per bin 
for both spectra. We show the fit obtained with a simple power-law model to 
highlight the iron line at $\sim$6.4~keV.}  
\label{fig:spectratatb} 
\end{figure}

All these results suggest that the drop in the count-rate noticed around $t$$\sim$23000~s 
was accompanied by the occurrence of a further spectral change in the X-ray emission from the source. 
The spectrum O$_{\rm 1}$ could not be described well by assuming a simple absorbed 
power-law model and the addition of a second more absorbed component (plus the iron line) 
provided a more reasonable fit to the data. In contrast, the spectrum O$_{\rm 2}$ could be well 
described by using a single relatively low absorbed power-law component. We show the two 
spectra O$_{\rm 1}$ and O$_{\rm 2}$ in Fig.~\ref{fig:spectratatb}, and 
discuss their interpretation further in Sect.~\ref{sec:discussion}.    

We note finally that the relatively low quality statistics of spectra A and O$_{\rm 2}$ and large errors 
affecting the values of their best-fit parameters prevent us from making a clear statement 
about any change in the spectral properties of the source 
before and after the occurrence of the flare. The \xmm\ data revealed, however, that the flux 
of the source before the onset of the event (time interval A) was a factor $\sim$3.5 
higher that that estimated during the last part of the observation (time interval O$_{\rm 2}$).

\subsection{Timing analysis}
\label{sec:xmmresultstiming} 

\begin{figure}
\centering 
\includegraphics[scale=0.43,angle=-90]{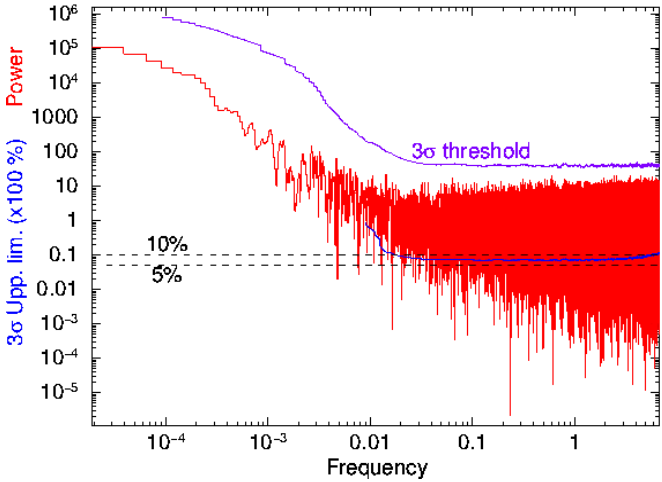}
\caption{Power spectrum produced using data from the entire \xmm\ observations 
(0.3-12 keV). The upper continuous line represents the power threshold for detection of 
periodicity at a  3~$\sigma$ c.l., according to the method described by \citet{israel96}. 
We also show in the bottom part of the plot a second continuous curve that represents 
the upper limit calculated with the same method, on the presence of pulsations from \ax\ 
as a function of the frequency. The two dashed lines represent the 5\% and 10\% upper limit 
levels on that curve. The most stringent upper limit we could provide with this method is 
6-10\% for frequencies in the range 0.02-6.8~Hz.}    
\label{fig:pulsed_frac} 
\end{figure}

\ax\ was reported in previous studies to emit pulsations at a period of $\sim$4.7~s 
\citep[][see also Sect.~\ref{sec:intro}]{bamba01,romano09}. 
We carried out an in-depth search for coherent modulations 
in data collected with both the Epic-pn and Epic-MOS cameras by extracting 
barycentered source and background event lists (we used the most accurately determined 
source position; see Sect.~\ref{sec:intro} and \ref{sec:xmmresultstiming}).   
We applied to the lists of barycentered photon arrival times in the 
0.3-12~keV energy range the power-spectrum search algorithm developed by \citet{israel96}. 
This method is optimized to search for periodicities in ``coloured'' power 
spectrum components and derive upper limits if no signal is detected. 
Pulsations were searched for by using both the photon arrival times of the 
entire \xmm\ observation and in the different time intervals selected for 
the spectral analysis. We did not find any significant indication of pulsations.  
As an example, Fig.~\ref{fig:pulsed_frac} shows the results obtained by applying 
the method of \citet{israel96} to the list of photon arrival times extracted from the entire 
\xmm\ observation. In Table~\ref{tab:xmmtiming}, we also report all the 3~$\sigma$ upper limits on the pulsed 
fractions\footnote{Here we define the pulsed fraction as the semi-amplitude of the sinusoid 
divided by the source average count rate, ($I_{\rm max}$-$I_{\rm min}$)/($I_{\rm max}$+$I_{\rm min}$).}  
and pulsation frequencies we investigated. We show in this table 
the upper limit estimated by using \xmm\ data accumulated during the entire 
observation (A+B+C+D+E+F+G+H+I+J+K+L+M+N+O), the flare event 
(including rise and decay, C+D+E+F+G+H), the rise of the flare (C+D), 
the top of the flare (E), the beginning of the flare decay (F+G+H), 
the last part of the flare decay (J+K+L), and the time interval in which the higher 
absorption was measured (M+N). This selection of intervals was carried out to investigate whether  
pulsations might have been present only during part of the event observed by \xmm.\  

For all the time intervals reported in Table~\ref{tab:xmmtiming}, we also performed additional  
searches of pulsations by separating the lists of photon time arrival times in the 
soft (0.3-4~keV) and hard (4-12~keV) energy bands and by combining strictly simultaneous 
EPIC-pn and MOS event lists (in this case, a common time resolution of 0.3-s was adopted). 
No significant improvement to the upper limits indicated in Table~\ref{tab:xmmtiming} 
could be obtained.  
\begin{table*}
\tiny
\centering
\caption{Three~$\sigma$ c.l. upper limits on the pulsed fraction of \ax\ (see notes below the table).}
\begin{tabular}{llll}
\hline
\hline
\noalign{\smallskip} 
\xmm\ (0.3-12 keV) & & & \\
\hline
\noalign{\smallskip} 
Time interval & \multicolumn{3}{c}{Instrument (time resolution)} \\
\hline
\noalign{\smallskip} 
& PN (73.4 ms) & MOS1 (0.3 s) & MOS2 (1.75 ms) \\
\hline
\noalign{\smallskip} 
A+B+C+D+E+F+G+H+I+J+K+L+M+N+O & 6-10\% (0.02-6.8 Hz) & 10-20\% (0.01-1.6 Hz) & 15-30\% (0.02-286 Hz) \\
\noalign{\smallskip}
C+D+E+F+G+H & 10-30\% (0.04-6.8 Hz) & 10-30\% (0.01-1.6 Hz) & 10-30\% (0.03-286 Hz) \\
\noalign{\smallskip}
C+D & 30-50\% (0.05-6.8 Hz) & 50-70\% (0.1-1.6 Hz) & 60-80\% (0.06-286 Hz) \\
\noalign{\smallskip}
E & 20-35\% (0.05-6.8 Hz) & 20-35\% (0.03-1.6 Hz) & 20-70\% (0.1-286 Hz) \\
\noalign{\smallskip}
F+G+H & 15-20\% (0.08-6.8 Hz) & 20-30\% (0.02-1.6 Hz) & 20-30\% (0.05-286 Hz) \\
\noalign{\smallskip}
J+K+L  & 10-20\% (0.02-6.8 Hz) & 20-40\% (0.01-1.6 Hz) & 35-60\% (0.02-286 Hz) \\
\noalign{\smallskip}
M+N & 40-60\% (0.02-6.8 Hz) & 75-100\% (0.02-1.6 Hz) & 80-100\% (0.02-286 Hz) \\
\noalign{\smallskip}
\hline
\hline
\noalign{\smallskip} 
\asca\ (0.7-10 keV) & & & \\
\hline
\noalign{\smallskip} 
 & GIS (0.5 s) & & \\
\hline
\noalign{\smallskip} 
Obs.1+Obs.2 & 30-45\% (0.01-1 Hz) & & \\
\noalign{\smallskip}
\hline
\hline
\end{tabular}
\tablefoot{The three~$\sigma$ upper limits on the pulsed fraction of \ax\ are 
determined for the three Epic cameras in the different time intervals reported in Table~\ref{tab:fit}. 
The results of a similar analysis carried out on the 
\asca\ data are also reported (see Sect.~\ref{sec:ascatiming} for the discussion of the ASCA data analysis).}
\label{tab:xmmtiming}
\end{table*} 
Even though for a number of time intervals considered 
in Table~\ref{tab:xmmtiming} the derived 3~$\sigma$ upper limit 
to the pulsed fraction was relatively high ($\sim$30-50\%), 
the entire exposure time of the \xmm\ observation provided  
relatively tight constraints on the presence of pulsations from \ax\ 
compared to those reported previously (see Sect.~\ref{sec:intro}). 
Motivated by these findings, we reanalyzed   
all the published \asca\ and \swift\ data for which 
detections of pulsations at $\sim$4.7~s were reported.

\section{ \asca\ data analysis}
\label{sec:ascadata} 

\ax\ was observed by ASCA for the first time on 1994 April 12 
(MJD 49454.677$-$49454.701), during the survey of the Scutum arm region 
(hereafter Obs.1). A follow-up observation was
carried out on 1999 October 3-4 (MJD 51454.252-51455.463, hereafter Obs.2).
On both occasions, \ax\ was outside the field of view (FOV) of the 
Solid-state Imaging Spectrometers (SIS, \citealp{sis}), but included in that of   
Gas Imaging Spectrometers (GIS, \citealp{gis}). 
The GISs were operated in the nominal PH mode with a time resolution of 62.5~ms for data taken 
in high bit-rate and 0.5~s for those taken in medium bit-rate. 
For our analysis, we used only GIS data obtained when the satellite was 
either outside the South Atlantic Anomaly and low cut-off rigidity 
regions ($>6$~GV), or when the target elevation angle was $>$5$^\circ$.
Particle events were removed based on rise-time discrimination 
with the {\sc gisclean} task \citep{gis}.
After this screening, the total available exposure times of Obs.1 and 2
were 2~ks and 36~ks, respectively. 
To improve the quality of the statistics, we combined data from the GIS-2 and GIS-3 detectors. 
\asca\ events were extracted from a circular region of 3~arcmin radius centered on the 
best known source position \citep{halpern04}.
The background was extracted from an annular region around the source position with inner radius
of 4.3~arcmin and outer radius of 6.8~arcmin. This annular region is free from field point
sources and is unaffected by  the extended diffuse emission from the nearby supernova remnant 
G$26.6-0.1$ \citep{bamba03}. The inner radius of the background extraction radius 
was chosen so as to avoid contamination due to the tail of the \asca\ PSF.  
Photon arrival times were all converted to the Solar System barycenter (SSB) 
with the task  {\sc timeconv} using the \chandra\ position reported by \citet{halpern04}. 
The log of the \asca\ observations of \ax\ is given in Table~\ref{log}. 
\begin{table}
\scriptsize
\centering
\caption{\asca\ observations log of \ax.\ }
\begin{tabular}{llrrr}
\hline
\hline
Instrument & Observation & Start time & Time elapsed & Exposure\\
\hline 
ASCA (1)&61009100 & 1994-12-04T15:48:13 & 4312.0  & 2112.0 \\
ASCA (2)&57026000 & 1999-10-03T05:47:32 & 65484.0 & 37373.6 \\ 
\hline
\end{tabular}
\label{log}
\end{table}

\subsection{Timing analysis}
\label{sec:ascatiming}

\begin{figure}
\centering 
\includegraphics[width=9cm,angle=0]{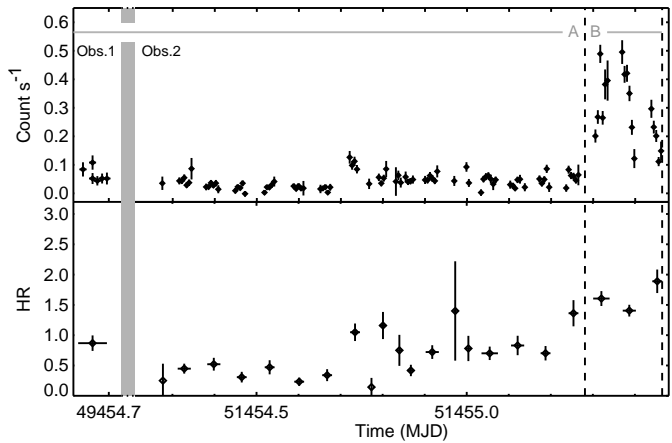}
\caption{{\it Top panel}: \asca\ light curve of \ax\ in the 0.7-10~keV energy band. 
{\it Bottom panel}: Hardness ratio of the 4-10 and 1-4~keV GIS light curves. 
The vertical dashed lines mark the flare episode starting at MJD~51455.3. 
The hashed region marks the separation between Obs.1 and Obs.2.
A and B denote the two data set selected for the timing analysis.}
\label{ascalc}
\end{figure}

In Fig.~\ref{ascalc}, we show the background-subtracted light curve (top panel) 
of \ax\ in the $0.7-10$ keV energy band. 
According to the results reported by \citet{bamba01}, we note that the source was 
characterized by a variable intensity, and that a 5~hr-long X-ray flare is visible 
at the end of Obs.2 (starting from MJD~51455.3). During this flare, the intensity of 
the source increased by a factor of $\sim$5. After $\sim$14 ks, \ax\ returned  
to an X-ray emission level comparable to that observed before the flare. 
The bottom panel of Fig.\ref{ascalc} shows the evolution of the hardness ratio computed by using 
the lightcurves extracted in the $0.7-2$~keV and $2-10$~keV energy bands. 
In Obs.2, the HR showed a remarkable increase (from $\sim$0.5 to $\sim$2) 
with the source intensity.

During the flare in Obs.2, \citet{bamba01} reported on the detection of coherent pulsations by carrying out  
a fast Fourier transform (FFT) of the barycentered photons-event arrival times selected in the $1.9-4.9$~keV 
energy range (energy channels $161-415$) and extracted from a $ 3^\prime$ circular region centered on 
the best known source position. An excess in the power spectrum (0.02--1 Hz) 
of $\sim$40 was visible at $\sim$0.211 Hz (their most accurately determined spin period was $P$=4.7394$\pm$0.0008~s, 
see also Sect.~\ref{sec:intro}). 
Taking into account the total  number of independent frequency trials, this excess 
corresponds to 4.4 (Gaussian) standard deviations. 

We revisited the data from Obs.2 performing a power spectrum analysis. 
The source events were collected during the flare 
with a time bin of 0.5 s (the time resolution of 
medium bit-rate ASCA data) and extracted from a $ 3^\prime$ radius circular 
region around the source. 
We investigated frequencies in the range 0.02-1~Hz with a resolution of 
3.7$\times$10$^{-5}$~Hz, first by using the entire energy range of the 
GIS (i.e. $0.7-10.0$~keV) and then only the events in the 
1.9-4.9~keV and 4.9-10.0~keV energy intervals. 

\begin{figure}
\centering 
\includegraphics[width=6.5cm,angle=-90]{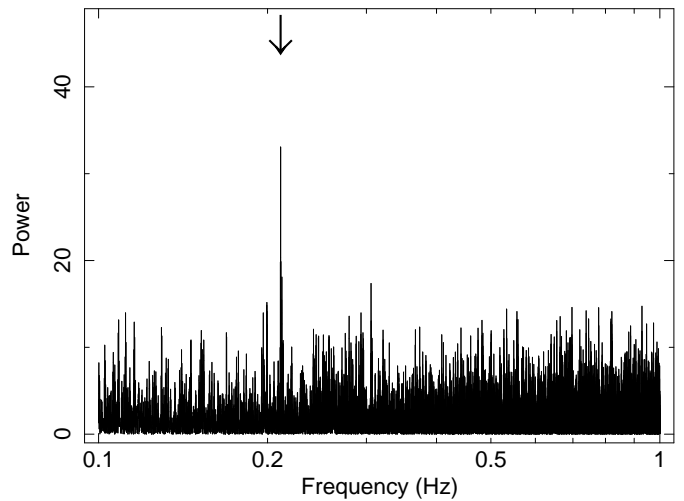}
\caption{ \asca\ power spectrum of \ax\ obtained by using events 
in the 1.9-4.9~keV energy range and a time resolution of 0.5~s. 
The arrow marks the apparent detection of pulsations at 0.211 Hz. 
As we discuss in the text, this excess in the power-spectrum 
is not statistically significant if the correct number of trials
is taken into account (see Sect.~\ref{sec:ascatiming}).} 
\label{ascapow}
\end{figure}

In the first and the latter cases, the power spectrum did not reveal any significant deviations 
from a statistically flat distribution. The corresponding estimated upper limit 
on the pulsed fraction of the source at different frequencies is reported 
in Table~\ref{tab:xmmtiming} (we used the method described in 
Sect.~\ref{sec:xmmresultstiming} in order to facilitate the comparison 
with the \xmm\ results). 
The power spectrum realized with the events in the energy interval 1.9-4.9~keV  
showed a peak at  $\sim$0.211~Hz with a power of $\sim$33 (see Fig~\ref{ascapow}). 
If all the inspected independent frequencies and the {\it ad-hoc} selection in energy 
are taken into account to estimate the total effective number of trials, this peak corresponds  
to a detection significance of $\lesssim$3 (Gaussian) standard deviations. 

A signiﬁcance of $\sim$4.4 (Gaussian) standard deviations 
for pulsations at $\sim$4.7~s could be obtained only by applying the FFT to 
the 1.9-4.9~keV flare events and using a time bin of 1s. 
Different choices of the time binning would led to a result
comparable to that obtained with our Fourier analysis (detection
signiﬁcance ~3 sigma).

Since pulsations in the ASCA data seems to be detected with
a suﬃciently high statistical significance only when a speciﬁc 
energy range is selected and only performing the analysis with the
FFT technique and an ad-hoc time bin of 1~s, we conclude that
this detection could be due to a statistical fluctuation.

\section{ \swift\ data analysis}
\label{sec:swiftdata}
 
Pulsations in the X-ray emission of \ax\ were also reported by \citet{sidoli08}. 
These authors used the three \swift\,/XRT observations 
ID.~00030988001, 00030988004, and 00030988005 (see Table~\ref{tab:swiftlog}).   
All data were collected in photon counting mode (PC, time resolution 2.5~s), and  
summed up to extract a single source photon event list and increase the statistics. 
By using a folding technique on this event list, 
\citet{sidoli08} found a peak in the periodogram at 4.70088$\pm$0.0004 s with 
a reduced $\chi^2$ of 6.3 (9 degree of freedom). 
\begin{figure}
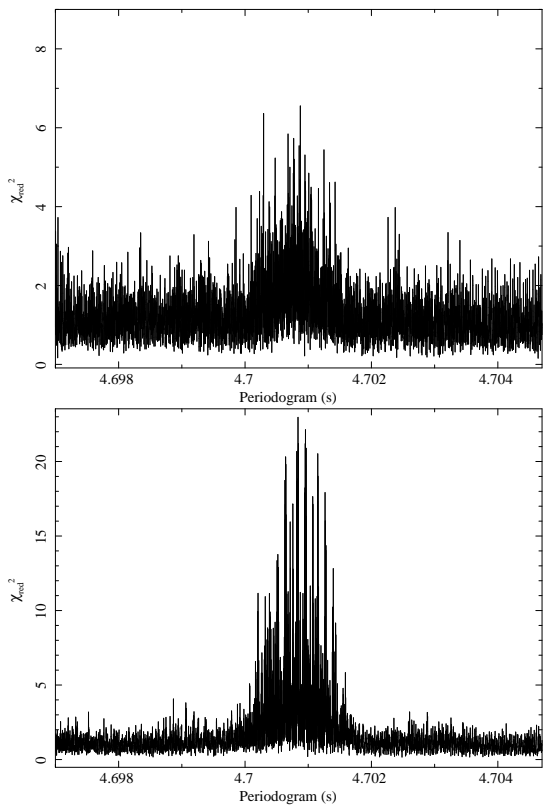

\centering 
\includegraphics[scale=0.3,angle=-90]{16726fig23.ps}
\includegraphics[scale=0.3,angle=-90]{16726fig24.ps}
\caption{Periodogram realized with the \swift\,XRT data by using 
the source (upper panel) and background (lower panel) event lists 
extracted by summing up data in the three observations ID.~00030988001, 
00030988004, and 00030988005.}
\label{fig:swiftperiodogram} 
\end{figure}

\subsection{Timing analysis} 

We reanalyzed the \swift\,XRT data of the above Swift-XRT observations 
to verify the detection of pulsations. 
We used the same \swift\ data analysis described in \citet{sidoli08}. 
XRT data were processed with standard procedures (xrtpipeline v. 0.11.6).  
Filtering and screening criteria were applied by using {\sc ftools} and only 
event grades 0-12 were considered. We extracted a source event list (194 events) by using a circular region 
of 29~arcsec centred on the best known position of \ax\ \citep{halpern04} and a background 
event list (1106 events) by using an annular region centered on the same  
position and with inner radius of 245~arcsec and outer radius of 630~arcsec. The radii of the background extraction region 
were chosen in order to avoid any contamination from the tail of the PSF\footnote{See also 
http://heasarc.gsfc.nasa.gov/docs/swift/analysis/xrt\_sw guide\_v1\_2.pdf.}.
The arrival times of the source and background events were all converted to the SSB with the task {\sc
barycorr}. 
\begin{table}
\tiny
\caption{ \swift\ observations log of \ax.\ } 
\begin{tabular}{@{}lllll@{}}
\hline
\hline
\noalign{\smallskip}
Obs. ID & Instr. & Start time & Stop time & Exp. \\
    &       &    (UTC)        &   (UTC)      &  (ks)   \\
\noalign{\smallskip}
\hline
\noalign{\smallskip}
00030988001 & XRT/PC & 2007-10-26 00:08 & 2007-10-26 06:45 & 1.4 \\
\noalign{\smallskip}
00030988004 & XRT/PC & 2007-11-05 09:08 & 2007-11-05 09:28 & 1.2 \\
\noalign{\smallskip}
00030988005 & XRT/PC & 2007-11-09 16:11 & 2007-11-09 16:32 & 1.3 \\
\noalign{\smallskip}
\hline
\end{tabular}
\tablefoot{Exp. indicates the effective exposure time of the observation.} \\
\label{tab:swiftlog}
\end{table} 
We applied the folding technique adopted by \citet{sidoli08} to the 
source and background event lists, to search for a period of $\sim$4.7~s.   
In the periodogram obtained from the analysis of source events (Fig.~\ref{fig:swiftperiodogram}, upper panel) 
the peak at 4.70088 ($\chi^2_{\rm red}$/d.o.f.=6.5/9) is visible.  
However, we noted that the same peak is also present in the periodogram obtained 
from the background event list (Fig.~\ref{fig:swiftperiodogram}, lower panel). In this case, 
we obtained $\chi^2_{\rm red}$=23. The difference in  the value of the 
$\chi^2_{\rm red}$ (a factor $\sim$4-5) is consistent with the different quality of the statistics content of the two data samples.
We thus argue that the detection of pulsations reported by \citet{sidoli08} was merely due to an 
instrumental effect present in the photon arrival times of the \swift\,/XRT data. 
Investigations of similar effects in other \swift\,/XRT data-sets are currently going on  
and will be reported in the relevant on-line pages\footnote{http://www.swift.ac.uk/xrtdigest.shtml.}.

\section{Discussion}
\label{sec:discussion}

During the 45~ks-long \xmm\ observation presented here, 
\ax\ underwent a bright X-ray flare lasting about 15~ks. The rise of the flare 
occurred 5~ks after the beginning of the observation, 
thus we were able to follow in detail  
the whole event to the return to quiescence.  
A time-resolved spectral analysis of the observation revealed that 
the source X-ray emission could be  
described well in the different selected time intervals 
by using an absorbed power-law model. 
The photon index of the power law remained virtually 
constant at least for the first 20~ks of the observation, 
but in this period the absorption column density underwent 
dramatic changes. We measured a significant rise in the 
$N_{\rm H}$ from $\sim$3 to 20$\times$10$^{22}$~cm$^{-2}$ across the transition 
from quiescence to the peak of the flare, 
followed by a sudden decrease to a value of 10$\times$10$^{22}$~cm$^{-2}$ which then 
remained stable for the following 4~ks. In the subsequent 11~ks,   
the $N_{\rm H}$ progressively increased to a value of $\sim$50$\times$10$^{22}$~cm$^{-2}$, 
and the spectral properties of the source changed dramatically 
during the last 23~ks of observation. In this time interval, we 
measured an abrupt flattening of the source power-law photon index and revealed  
the appearance of a prominent emission line at $\sim$6.4~keV. 
Similar behavior is usually observed during the occurrence of X-ray eclipses 
in some of the sgHMXBs and in only one SFXT \citep[IGR\,J16479-4514, see e.g.,]
[and references therein]{bozzo08}. 
When the NSs in one of these systems is obscured by its supergiant companion,  
the X-ray emission produced by the accretion process close to the surface of the compact object 
is progressively absorbed in the photosphere of the supergiant star. Only 
the X-rays reflected by matter in the stellar wind remain  
visible to the observer and a fluorescence emission line at $\sim$6.4~keV 
becomes very prominent.  

In the present case, that \ax\ underwent 
a bright X-ray flare after which the source turned off sharply because of an X-ray eclipse 
appears to be unlikely. 
Below we discuss an alternative scenario, in which the entire event observed 
by \xmm\ is interpreted as the accretion of a massive clump onto the 
compact object hosted in \ax.\   
\begin{figure*}
\centering 
\includegraphics[scale=0.18]{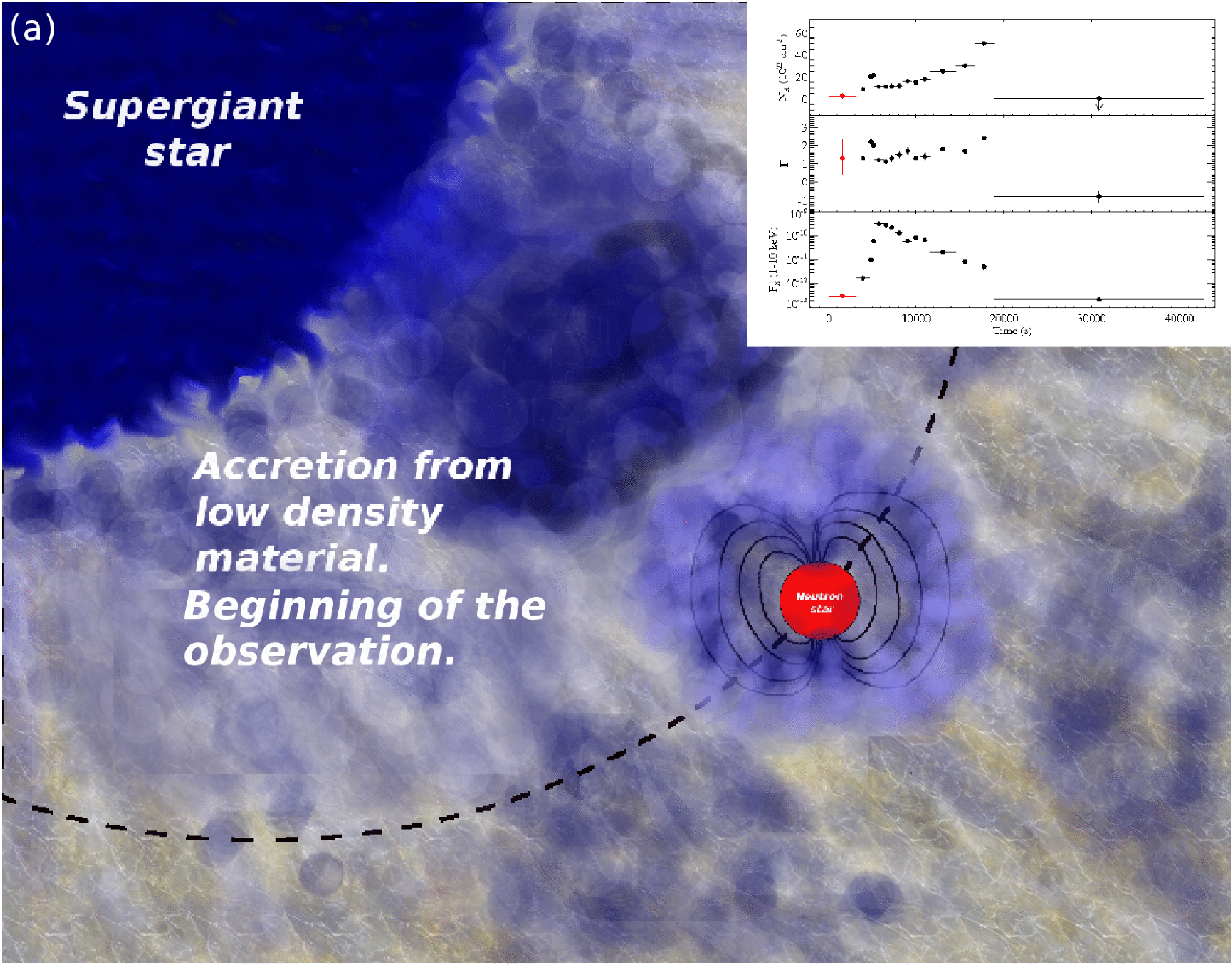}
\includegraphics[scale=0.18]{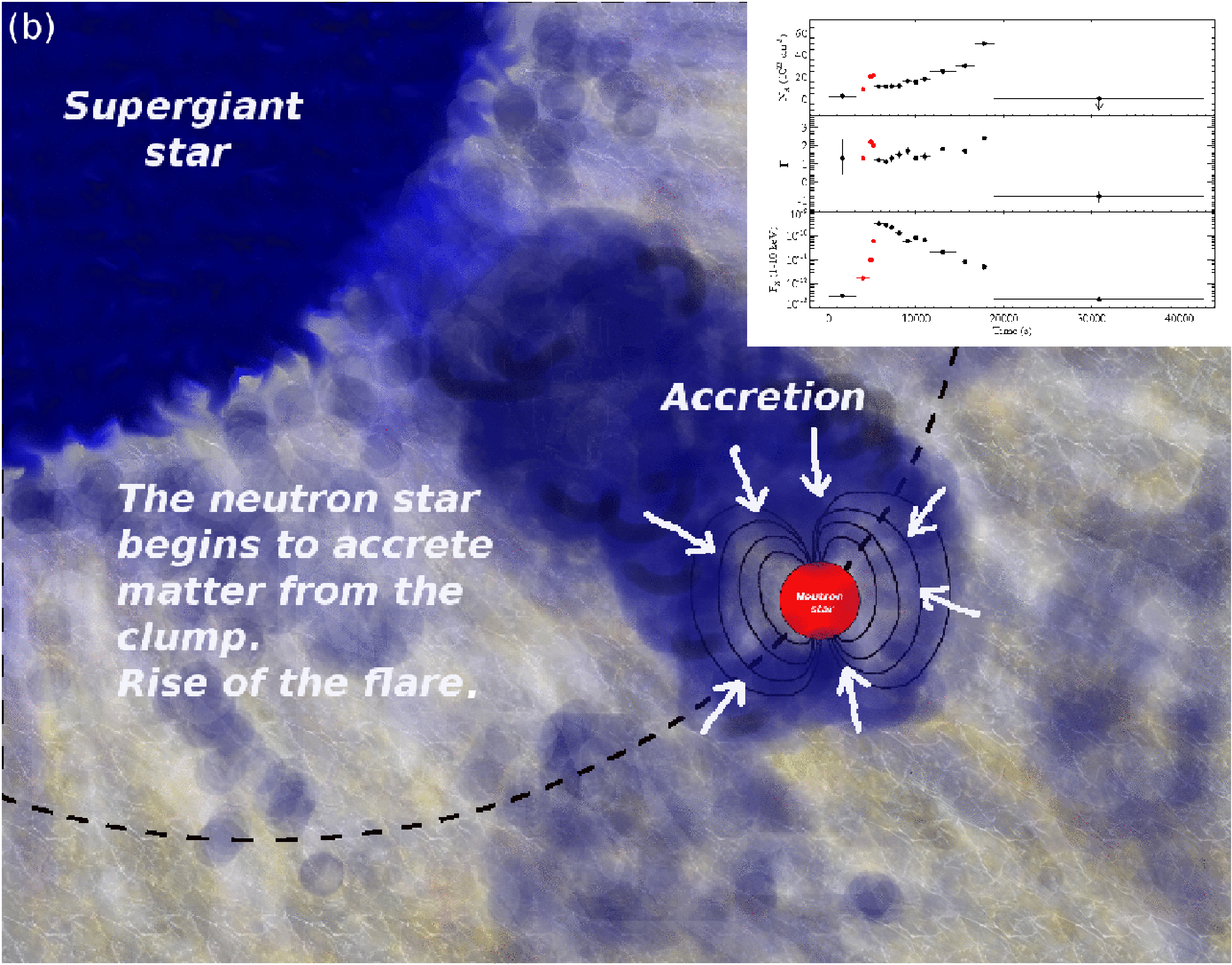}
\includegraphics[scale=0.18]{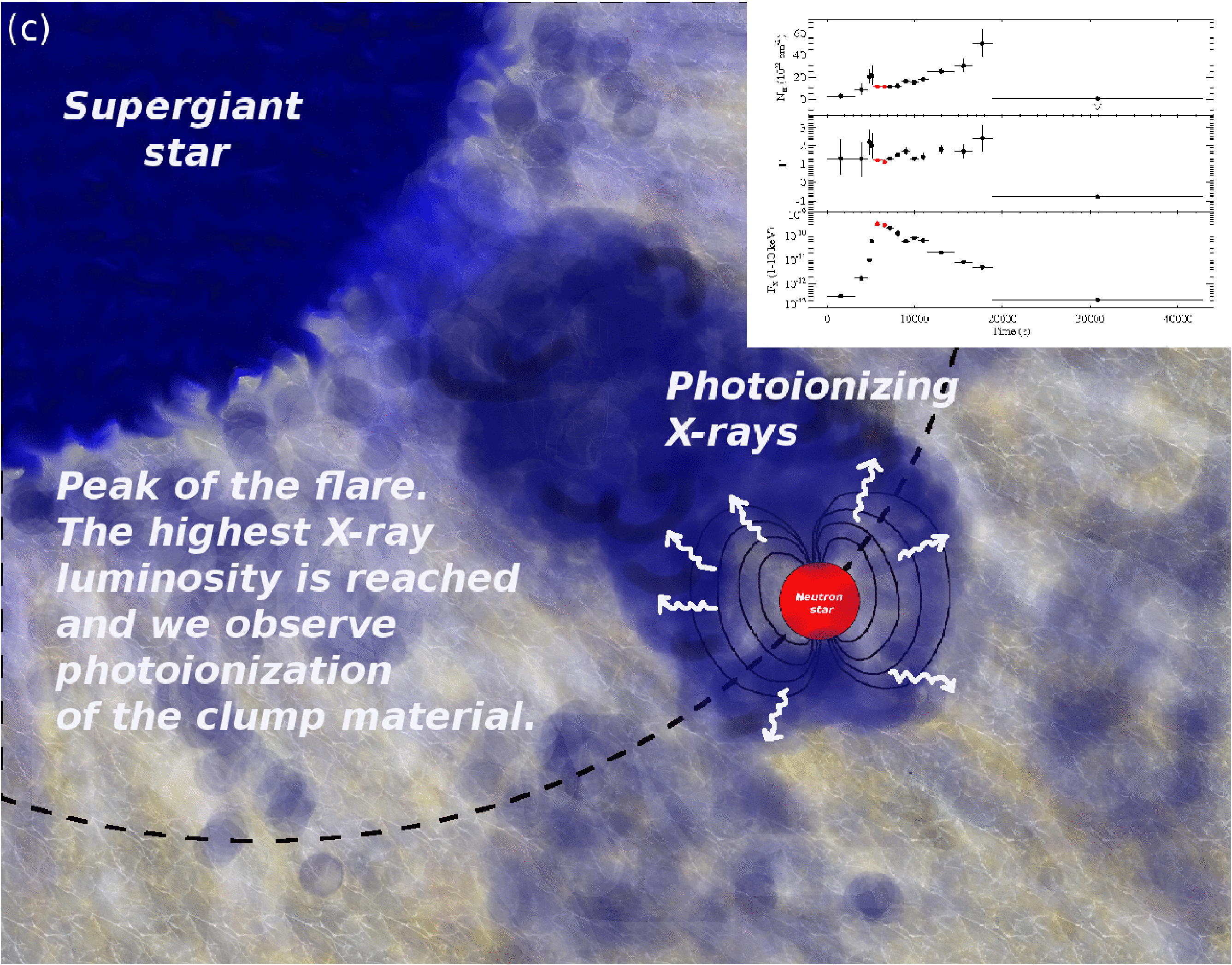}
\includegraphics[scale=0.18]{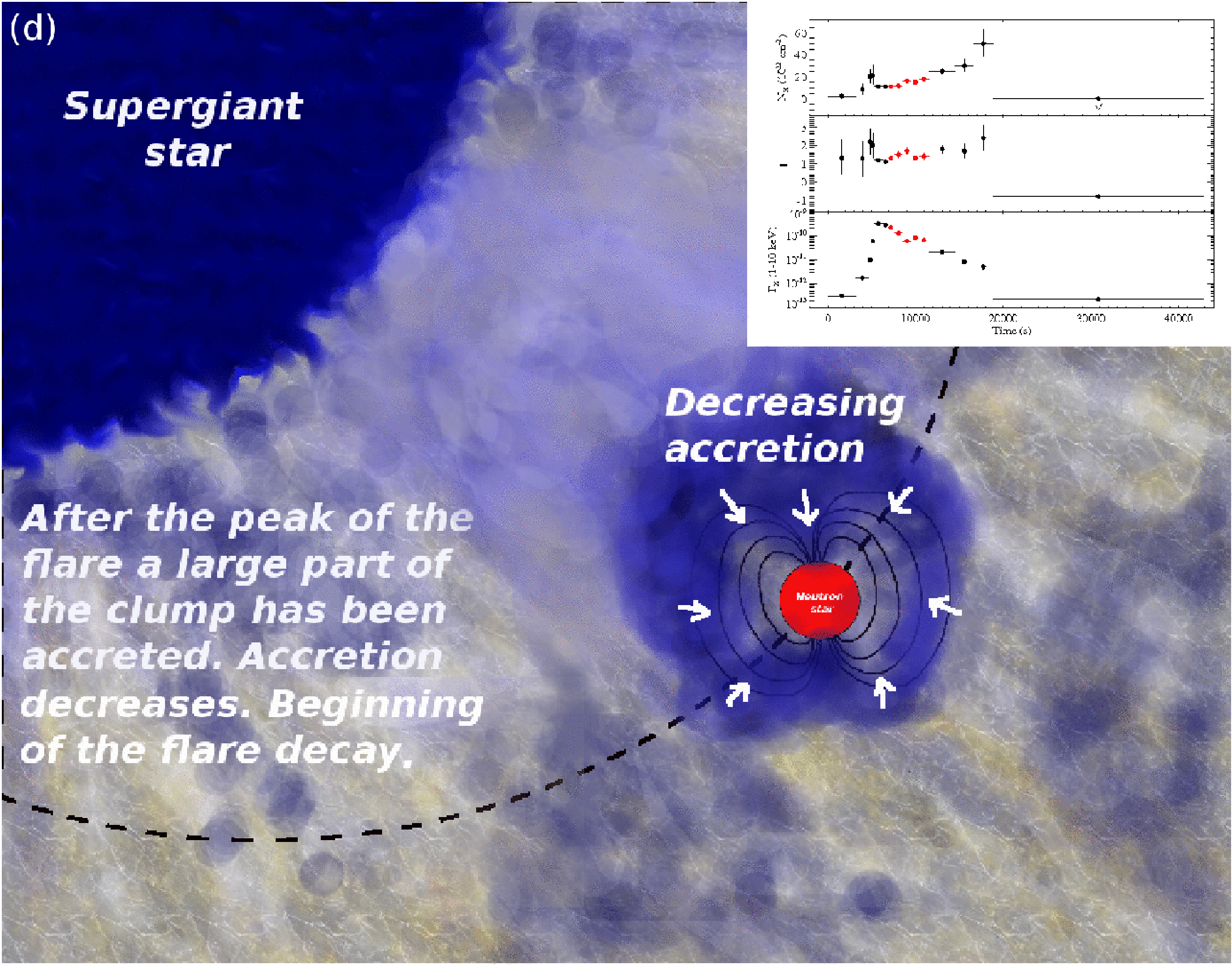}
\includegraphics[scale=0.18]{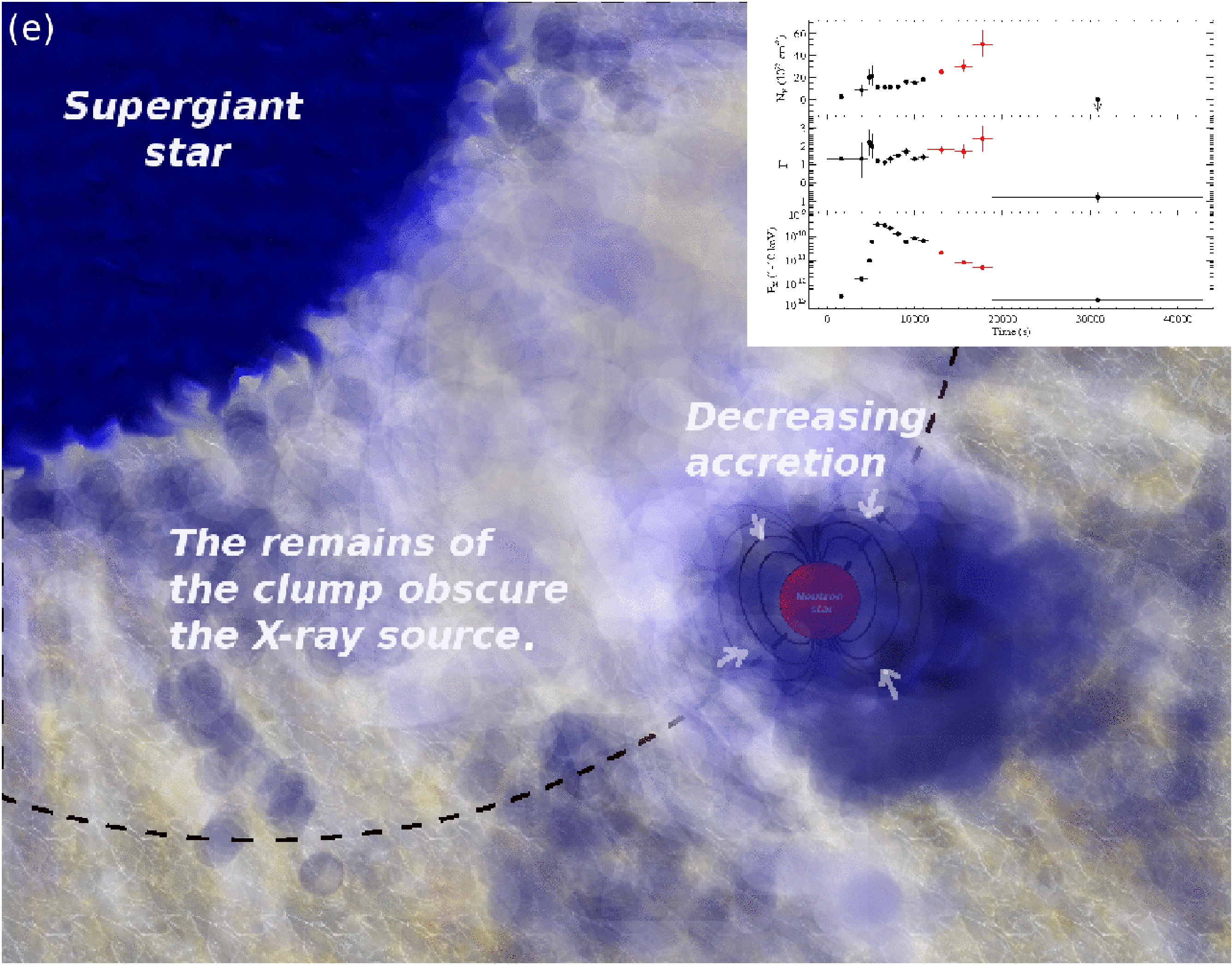}
\includegraphics[scale=0.18]{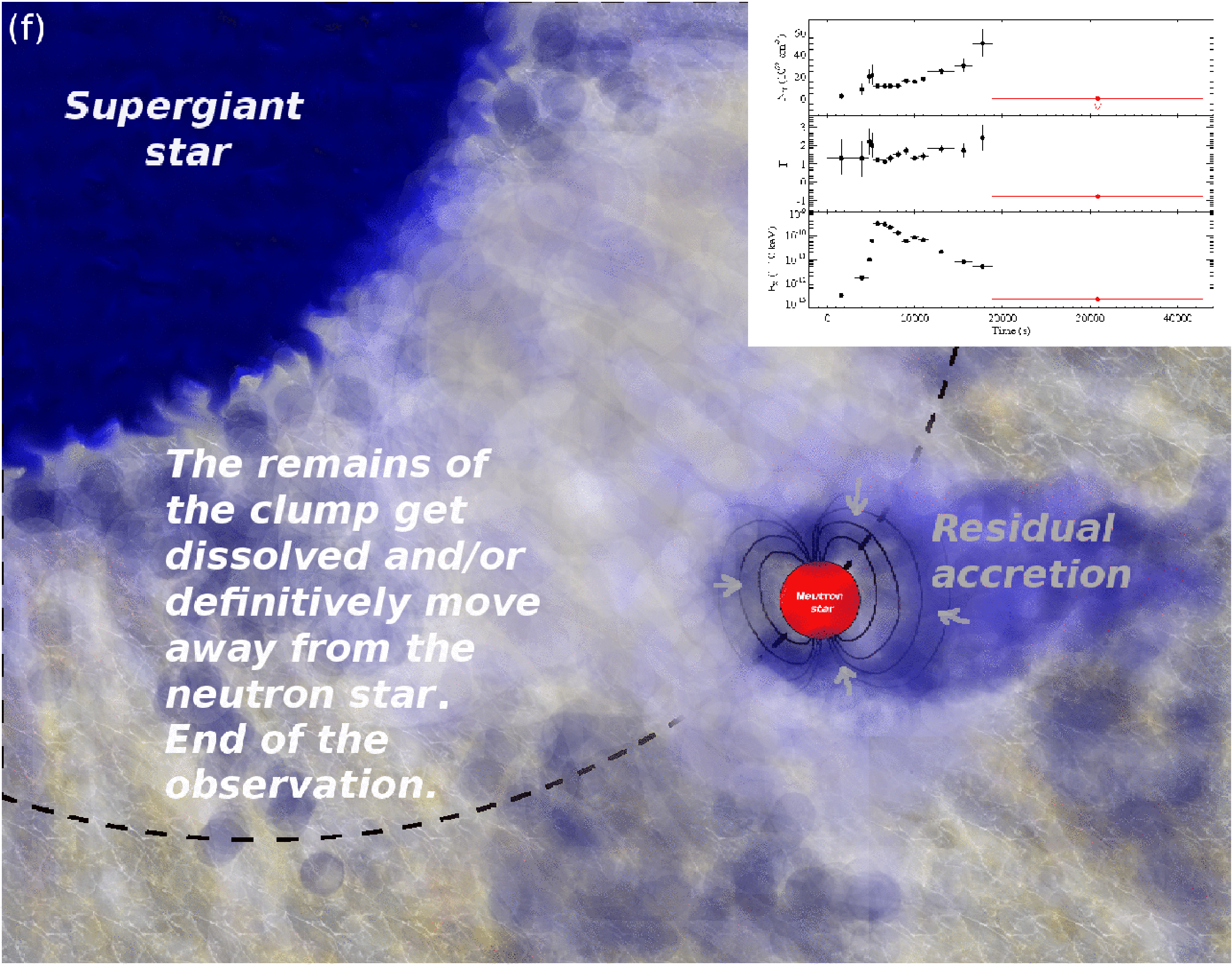}
\caption{An artist's view of the ``ingestion of a clump'' described in 
Sect.~\ref{sec:ingestion}. Different elements in the figure are not represented 
to scale. The figures illustrate qualitatively the main phases of the event 
in chronological order from the top left to the bottom right. In each figure, we plot 
in the top right corner the \xmm\ lightcurve of the event already shown in Fig.~\ref{fig:parameters} 
and highlight with a red color the data points in the plot that 
correspond to the physical scenario described in the picture. From top left to 
bottom right: (a) at the beginning of the \xmm\ observation the NS is accreting from a 
low density material, such that the observed X-ray luminosity is relatively low 
($\sim$4$\times$10$^{32}$~erg$s^{-1}$); (b) the NS encounters the clump and \xmm\ observes 
a rapid rise in the X-ray flux from the source; (c) the higher X-ray luminosity 
($\sim$4$\times$10$^{35}$~erg$s^{-1}$) causes the photoionization of the surrounding material 
and a decrease in $N_{\rm H}$; (d) accretion decreases and the source undergoes a rapid 
decay in the X-ray flux; (e) the remains of the clump move in front of the 
NS along the observer line of sight and obscure the X-ray source; (f) the remains of the clump 
move away and the source returns to its quiescent state 
($\sim$10$^{32}$~erg$s^{-1}$, some residual accretion might still take place).} 
\label{fig:clump} 
\end{figure*}

\subsection{Ingestion of a massive clump}
\label{sec:ingestion} 
 
The idea that dense ``clumps'' of matter might be generated in the wind of supergiant stars  
as a consequence of different magnetohydrodynamic instabilities has found some support in the 
results of several numerical simulations \citep[][and references therein]{runacres02,oskinova07} 
and in observations of isolated OB stars \citep[see e.g.,][]{eversberg98,martins10}. 
Even though the physical parameters of these clumps 
(i.e., mass, radius and density) are still largely unknown because of   
poorly estimated theoretical and observational parameters, a clumpy wind scenario is the 
founding hypothesis of most models developed so far to interpret the X-ray variability of the 
SFXTs \citep[][and references therein]{zand05,walter07,bozzo08b}.
All these models predict that fast X-ray flares can be produced by 
the sporadic capture and accretion of one of these clumps onto the NSs hosted in the 
SFXTs. The role of these dense clumps distributed all around the NS orbit is two-fold. Clumps
simply passing in front of the X-ray source cause dimming or even obscuration, and display the
signatures of photoelectric absorption, at least during the ingress and egress stages. In addition to these
phenomena, clumps that lead to increased accretion also give rise to large variations in the X-ray
luminosity. To reach the required X-ray luminosity level, different densities of the clumps 
might be required, depending also on the strength of the NS magnetic field and 
the value of its spin period. Both the rotation of the NS and the extent  
of its magnetosphere can act as gates that halt most of the accretion during quiescence 
and release it during the outburst, thus significantly altering the total dynamic range 
of the X-ray luminosity \citep{grebenev07,bozzo08b}. 

In the case of \ax,\ the entire flaring episode observed 
by \xmm\ might be due to the accretion of a single massive clump. 

\begin{itemize}
\item $t$=0-3200~s (time interval A): {\it Pre-flare quiescence.} 
At the beginning of the \xmm\ observation, the source was caught at a relatively low X-ray flux 
(3.2$\times$10$^{-13}$~erg~cm$^{-2}$~s$^{-1}$). For 
a source distance of 3~kpc, this would translate into an X-ray luminosity of 3.5$\times$10$^{32}$~erg/cm$^{2}$/s, 
i.e. a value typical of SFXTs in quiescence. The spectral parameters, $N_{\rm H}$, $\Gamma$, that we measured in 
this time interval (spectrum A in Table~\ref{tab:fit}) are compatible with those inferred for the 
``very low state'' reported by  \citet{romano09}. The relatively low flux and absorption column density 
measured in this state with respect to the other average X-ray emission states (``medium'' and ``high'') suggests that the NS 
at the beginning of the \xmm\ observation might have been located in a tenuous region of the stellar 
wind such that accretion was taking place only at a very low level or that it was 
inhibited by magnetic and/or centrifugal barriers. 

\item $t$=3200-5220~s (time intervals B, C, and D): {\it Rise of the flare}. 
Around $t$=3200~s, the \xmm\ lightcurve showed a remarkable rapid increase in the count-rate from \ax.\ This  
is firstly noticeable in the hard energy band (4-12~keV, see Fig.~\ref{fig:rise}), 
thus indicating a high level of absorption. During the rise of the flare, we measured a significant increase in   
$N_{\rm H}$, reaching a value of $\sim$2$\times$10$^{23}$~cm$^{-2}$ at the top of the flare ($t$$\simeq$5200~s). At this stage, 
the source X-ray flux increased to 3.3$\times$10$^{-10}$~erg~cm$^{-2}$~s$^{-1}$, corresponding to an X-ray luminosity of 
3.6$\times$10$^{35}$~erg/s. A similar simultaneous increase in the X-ray luminosity and absorption column density 
can be well understood by assuming that a massive clump approached the NS magnetosphere and filled the  
immediate surroundings with high density material. The increase in the local density can easily lead to a significant 
increase in the $N_{\rm H}$ along our line of sight and eventually compress the NS 
magnetosphere to open the magnetic and/or centrifugal barrier and permit direct accretion onto the compact object. 
In this interpretation, we can use the observational results in Table~\ref{tab:fit} to estimate 
the physical properties of the clump. 

In wind accretion theory, the typical time scale on which matter is accreted and reaches the surface 
of the NS from the magnetospheric boundary is of the order of the local free-fall time, i.e. hundreds of 
seconds \citep[see e.g.,][]{bozzo08b}. 
The duration of the flare observed from \ax\ is $\sim$15~ks, thus this time is likely to be linked 
to the radial extent of the clump. Therefore, we can estimate 
\begin{equation}
R_{\rm cl} \simeq 1/2 v_{\rm w} \Delta t_{\rm flare} = 8\times10^{11} v_{\rm w8}~{\rm cm},  
\end{equation} 
where $R_{cl}$ is the radius of the clump and $v_{\rm w8}$ is the relative velocity between the clump and the NS 
in units of 10$^{8}$~cm/s \citep[we neglect here the orbital velocity 
of the NS and consider a spherical clump moving with the same velocity as the 
surrounding stellar material;][]{lepine08}. To a first order approximation, only the material that 
falls inside the accretion radius, $R_{\rm acc}$, of the NS is accreted, and thus 
\begin{equation}
M_{\rm cl}= (R_{\rm cl}/R_{\rm acc})^2 M_{\rm acc}, 
\end{equation}
where $R_{\rm acc}$=2$G$$M_{\rm NS}$/$v_{\rm w}^2$, 
$M_{\rm NS}$=1.4$M_{\odot}$ is the 
mass of the NS, and $M_{\rm acc}$ is the total mass accreted during the flare. 
The latter can be estimated by integrating the X-ray flux measured during the flare as a function of time 
in Fig.~\ref{fig:parameters} and using the relations $F_{unabs}$=$L_{\rm X}$/(4$\pi$$d^2$) and 
$L_{\rm X}$=$G$$M_{\rm NS}$$\dot{M}_{\rm acc}$/$R_{\rm NS}$ (where $d$ is the source distance, $R_{\rm NS}$=10$^6$~cm 
is the NS radius and $\dot{M}_{\rm acc}$ is the mass accretion rate). 
We found $M_{\rm acc}$=1.5$\times$10$^{19}$$d_{\rm 3kpc}$~g (where $d_{\rm 3kpc}$ is the 
source distance in units of 3~kpc) and thus 
\begin{equation}
M_{\rm cl}= 1.4\times10^{22} v_{\rm w8}^6  d_{\rm 3kpc}^2~{\rm g},  
\end{equation}
which clearly has a strong dependence on the wind velocity. 
Another equation that can be used as a cross-check of these results is that relating the mass 
and radius of the clump to the expected absorption column density caused by its presence in the 
vicinity of the NS, i.e. 
\begin{equation}
N_{\rm H}\simeq M_{\rm cl}/(R_{\rm cl}^2 m_{\rm p})= 1.3\times10^{22} v_{\rm w8}^4 d_{\rm 3kpc}^2 {\rm cm^{-2}}. 
\label{eq:nh} 
\end{equation}
As the $N_{\rm H}$ measured by the spectral analysis at the top of the flare is $\sim$2$\times$10$^{23}$~cm$^{-2}$, 
Eq.~\ref{eq:nh} implies a wind velocity slightly lower than 10$^8$~cm/s. 
We note that the estimated values of $M_{\rm cl}$ and $R_{\rm cl}$ are in qualitative agreement with those 
expected according to the clumpy wind model developed by \citet{ducci09}.  

\item $t$=5220-6320~s (time interval E): {\it Top of the flare}. 
The results of the spectral analysis obtained during this time interval can also be clearly explained within 
the scenario depicted above. The time-resolved spectral analysis carried out in Sect.~\ref{sec:xmmresultsspectra} 
showed that the initial rise in the $N_{\rm H}$, interpreted above as being due to a massive clump approaching the NS, 
was followed by a sudden decrease in the absorption column density (by a factor of $\sim$2) when  
the system reached the highest luminosity. According to the calculation of \citet{krolik84}, a similar drop in  
$N_{\rm H}$ is to be expected because of the heating and photoionization of the clump material 
by the higher X-ray flux.  In particular, a major change in the opacity of the material with respect to the X-ray 
photons is expected when the so-called 
ionization parameter $\Xi$$\simeq$10. The latter is defined as 
\begin{equation}
\Xi = \frac{L_{\rm X}}{4 \pi r^2 N k T} \simeq 8 v_{\rm w8}^{-5} d_{\rm 3kpc}^{-2} L_{\rm 35} T_{5}^{-1},  
\end{equation}
where $L_{\rm 35}$=$L_{\rm X}$/(10$^{35}$~erg/s) is the luminosity coming from the X-ray source, $r$$\simeq$$R_{\rm cl}$ 
is the distance of the material from the source, and $T$$\simeq$10$^5$~K is the temperature of the clump \citep{ducci09}.  
We therefore expect $\Xi$ to be close to the critical value for a luminosity 
similar to that reached by \ax\ at the peak of the flare. 
During this time interval, we also detected a significant iron line with a centroid energy 
of 6.56$^{+0.06}_{-0.05}$~keV. This is somewhat higher than the Fe $K_{\alpha}$ line of neutral iron  
($\sim$6.4~keV), thus suggests that the X-ray flux produced by accretion onto the NS  
partly ionized the clump matter \citep[according to][the centroid energy we measured at the top of the flare 
requires the presence of iron ions with ionization stages higher than FeXXI]{kallman04}. 

\item $t$=6320-9520~s (time intervals F, G, H, and I): {\it Beginning of the flare decay}. 
During this interval, the source X-ray flux was beginning to decrease.  
When it dropped below $\sim$6$\times$10$^{-11}$~erg~cm$^{-2}$~s$^{-1}$, 
a new rise in the $N_{\rm H}$ is measured (spectrum I). 
This is to be expected according to our interpretation where the X-ray flux 
decreases below the level required for photoionising the clump material. 
A further possible indication of the change in the ionization stage 
of the material around the NS can be deduced by the energy of the iron 
line detected during interval I. This is somehow lower than that 
measured at the top of the flare and close to the value expected in case 
of neutral iron (see Fig.~\ref{fig:lines}). 

An enigmatic finding that emerged from the analysis of the \xmm\ data during this time interval 
is the drop in the source count-rate visible in Fig.~\ref{fig:total_lcurve} around $t$=8500-9500. 
As we discussed in Sect.~\ref{sec:xmmresultsspectra}, during this interval no particular spectral change 
was observed in the X-ray emission (we note only a marginal increase in the $N_{\rm H}$ 
with respect to the spectra extracted from the nearby time intervals and detected a significant 
iron line at $\sim$6.4~keV, see Table~\ref{tab:fit}). 
A closer inspection of the \xmm\ lightcurve of the source around $t$=8500-9500 revealed 
that the apparent drop in count-rate occurred in both the low (0.3-4~keV) and hard (4-12~keV) energy bands  
and that several small flares took place during this period (see Fig.~\ref{fig:hole}). We suggest  
here that a qualitatively similar variations in the X-ray flux from the source might  be expected because of some 
instabilities taking place close to the NS magnetosphere \citep[e.g., when the decreasing accretion 
rate of material from the clump is close to the threshold for the opening/closing of the centrifugal 
and/or magnetic barrier][]{bozzo08b}. Alternatively, it can be related to some small structures 
in the stellar wind located close or within  
the clump itself.    

\item $t$=9520-42840~s (time intervals J, K, L, M, N, and O): {\it Late stages of the flare decay}. 
The spectral analysis revealed a progressive  
increase in the absorption column density, that reached a value of 
$N_{\rm H}$$\simeq$50$\times$10$^{22}$~cm$^{-2}$ in the time interval N. 
Even though an increase of the $N_{\rm H}$ is still expected 
at these times because of the lower ionizing X-ray flux, the value above   
is significantly higher than that measured when the source reached the peak of the X-ray flux.
This result can be explained by assuming that, after the occurrence of the flare, the clump was located 
in front of the NS along our line of 
sight to the source. This would agree with the dramatic change in the X-ray spectrum of the 
source observed at the end of the flare during interval O ($t$=18870-42840~s). 
We suggested in Sect.~\ref{sec:xmmresultsspectra} that a reasonable fit to this spectrum 
could have been obtained with a model comprising two power-law components with different 
absorptions plus an iron line at 6.4~keV. This model is usually adopted to interpret 
the X-ray emission of sgHMXBs in eclipse \citep[see e.g,][for a review]{meer05,torreion10}, 
but in the case of \ax,\ we assume that the obscuration of the NS was caused by a massive clump rather 
than by its supergiant companion. 
The more absorbed power law represents 
the X-ray emission caused by the accretion onto the NS that is strongly extinguished by the presence of the clump.  
The less absorbed power-law component is introduced to take into account the scattered emission in the 
stellar wind material spread all around the binary system, and might also comprise the X-ray emission 
produced by the shocks in the wind itself \citep[see e.g.,][and references therein]{bozzo10}.  
This emission is unaffected by the presence of the clump and is thus only seen through an absorption column density 
that is compatible with the Galactic value. The iron line is caused by X-ray irradiation of cold iron 
in the wind of the supergiant star \citep[see, e.g.][]{kallman04}, and its relatively high $EW$ 
also support the idea that most of the continuum emission is suppressed by the obscuration of the clump. 
However, if we consider that the obscuration of the X-ray source
is taking place during the entire time interval O, some {\it ad hoc} assumptions on the system 
geometry would probably be required to explain how a clump with the physical dimensions 
estimated earlier in this section and moving with a velocity of the order of $\sim$10$^8$~cm/s can hide the NS 
for about $\sim$20 ks. We suggest instead that the obscuration 
occurred only during the first $\sim$3~ks of the time interval O (i.e. time interval O$_{\rm 1}$). 
Once the clump moved away, the accretion onto the NS decreased substantially and the X-ray emission detected during the 
last $\sim$20~ks of observation (time interval O$_{\rm 2}$) was partly due to the residual accretion and partly to  
the scattering  of these X-rays in the wind of the supergiant star and the shocks occurring within the wind itself 
\citep[see e.g.,][and references therein]{bozzo10}. 
Even though the statistics of the data accumulated during 
the time intervals O$_{\rm 1}$ and O$_{\rm 2}$ was of too low quality to enable us to draw a firm conclusion, 
the analysis of these spectra 
carried out in Sect.~\ref{sec:xmmdata} is consistent with this idea. 
We showed that a fit to the O$_{\rm 1}$ spectrum required both the presence of the highly 
absorbed and the less absorbed power-law 
component, whereas the spectrum O$_{\rm 2}$ could be described well using only the less absorbed power-law. 
However, given the very poor spectral information available during the time interval O$_{\rm 2}$, we cannot presently 
rule out that after the clump moved away, the obscuration of the NS by its supergiant companion
contributed to reduce the flux measured at this time 
($\sim$9$\times$10$^{-14}$~erg~cm$^{-2}$~s$^{-1}$, 1-10 keV, i.e a factor of $\sim$3.5 lower 
than that measured in the interval A, see Table~\ref{tab:fit}). 

As a final remark, we also note that an iron line with a centroid energy and an $EW$ compatible with that measured 
during the time interval I and E by \xmm\ was detected from \ax\ during the \asca\ observations carried 
out in 1994 and 1999 (see Sect.~\ref{sec:intro}). On those occasions, 
the line was clearly visible in the spectra of the source extracted in the flux range  
(0.1-1.0)$\times$10$^{-11}$~erg~cm$^{-2}$~s$^{-1}$ (2-10~keV), i.e. significantly different from  
those corresponding to the \xmm\ detections (see Table~\ref{tab:fit}). No convincing  
indication of the presence of an iron line could be found in the \xmm\ spectra extracted 
at similar flux levels (intervals L, M, N; see also Sect.~\ref{sec:xmmresultsspectra}). As the centroid energy of the 
iron line, together with its $EW$ and normalization, provides information on the ionization state of stellar wind material 
and the amount of material around the NS, these results support the idea that the conditions of the stellar winds in the SFXT sources 
can change significantly during the NS orbits \citep[see e.g.,][]{walter07}. 
\end{itemize}

\subsection{\ax:\ another SFXT hosting a slow-spinning NS?} 
\label{sec:spin} 

The timing analysis of the \xmm\ data did not reveal any 
pulsations at 4.7~s, which is the periodicity detected during previously 
\asca\ and \swift\ observations of \ax\ (see Sect.~\ref{sec:intro}).  
The relatively tight upper limits on the pulsed fraction 
derived from the \xmm\ data (see Table~\ref{tab:xmmtiming}), 
convinced us to reanalyze the archival \asca\ and \swift\ data where pulsations 
were detected. In the first case, our results revealed that the detection might 
have been due to a statistical fluctuation, while in the second case the detection was most likely 
related to an instrumental problem. In addition the analysis of the latest \swift\ observations 
performed during the outburst of the source that occurred on 2010 June 5 
could not confirm the presence of pulsations \citep{romano10b}.  
These results suggest that the NS in \ax\ might not be pulsating 
at a period of 4.7~s. 

Among the $\sim$15 sources in the SFXT class, so far only IGR\,J11215-5952, 
IGR\,J18483-0311, and IGR\,J16465-4507, displayed unambiguously coherent  
pulsations at $\sim$186.78~s, $\sim$21~s, and $\sim$228~s, respectively 
\citep{sidoli07,walter07,giunta09}.  
However, IGR\,J11215-5952  proved to be a very peculiar source, as it is the only 
SFXT displaying regularly periodic outbursts connected with the NS passage  
at the periastron of the system, while IGR\,J18483-0311 and 
IGR\,J16465-4507 were classified as ``intermediate SFXTs'' due to their 
somewhat longer outbursts (up to a few days as opposed to a few hours) and  
smaller dynamic range in the X-ray flux variation \citep[$\lesssim$10$^3$, see e.g.]
[]{rahoui08,romano09b,laparola10,clark10}. All SFXTs 
displaying a very large dynamic range in the X-ray luminosity ($\gtrsim$10$^4$-10$^5$), 
including \ax,\ seem instead to be ``non-pulsating'' sources  
\citep[see the cases of, e.g. XTE\,J1739-302 and 
IGR\,J08408-4503 and IGR\,J17544-2619][]{bozzo10,rampy09}.  
In some of these cases, pulsations were searched deeply up to periods of several 
hundred seconds, and no evidence for them was found 
\citep{smith06,bozzo09,bozzo10}. These non-detections 
are still consistent with the possibility that these 
SFXT sources might host NSs with very long spin periods 
($\gtrsim$1000~s) and that their extreme X-ray variability could  
be related to the centrifugal and magnetic gating mechanism 
\citep{bozzo08b}.

\section{Conclusion}
\label{sec:conclusion}
 
We have reported on the detection with \xmm\ 
of a relatively bright flare from the SFXT source \ax.\ 
The lightcurve and spectral analysis presented in 
Sect.~\ref{sec:xmmdata}, provided several indications 
that the event was entirely due to the accretion of a massive 
clump onto the NS hosted in this system. 
Even thought the occurrence of an eclipse by the companion 
star toward the end of the observation could not be 
completely ruled out (see Sect.~\ref{sec:discussion}), 
we provided convincing evidence in favor of the 
``ingestion of a clump'' scenario. We have shown that: 
\begin{itemize}
\item the rise of the flare was followed by a sudden increase in the 
absorption column density in the direction of the source. This is expected 
if the event is triggered by the presence of a massive clump approaching the NS. 
\item at the peak of the flare, a significant drop in the absorption 
column density occurred; this was ascribed to the effect of the X-ray photoionization 
caused by the higher X-ray luminosity of the source (this is also supported 
by the evidence of a change in energy of the iron line centroid with 
the source luminosity). 
\item the gradual increase in the absorption column density in the direction of the 
source during the decay from the top of flare, together with the detection of a prominent 
iron line in the last part of the \xmm\ observation, could be interpreted as implying 
that the remains of the clump first covered the X-ray source for a few 
ks ($\sim$3~ks) and then moved away from the NS. The residual X-ray emission observed 
in the last $\sim$17~ks of the observation might be due to some residual accretion 
taking place onto the NS.  
\end{itemize}
In accordance with this interpretation, we have provided a rough estimate of the mass and radius 
of the clump by assuming a spherical shape. We note that, because of our poor knowledge of the 
properties of the binary system \ax\ and the lack of a clear confirmation of its spin 
period (see Sect.~\ref{sec:spin}), it was not possible to take into account in these calculations 
the effect of the NS magnetic field according to the model developed by \citet{bozzo08}. 
The lack of clear detection of relatively short spin periods in all the 
most extreme SFXT sources, including \ax,\ is still consistent with the possibility 
that these object might host NSs with very long spin periods 
($\gtrsim$1000~s) and that their extreme X-ray variability could  
be partly related to the centrifugal and magnetic gating mechanisms  
\citep{bozzo08b}.

\begin{acknowledgements} 
EB thanks the Swiss Society for Astronomy and 
Astrophysics for granting the on-going collaboration 
between the ISDC (Geneva, Switzerland), the ISSI 
(Bern, Switzerland) and the INAF-OAR (Rome, Italy). 
LS acknowledges financial support from ASI. 
We thank the anonymous referee for useful comments. 
The results presented in this paper are based on observations obtained 
with XMM-Newton, an ESA science mission with instruments and contributions 
directly funded by ESA Member States and NASA.  
\end{acknowledgements}

\bibliographystyle{aa}
\bibliography{16726}

\end{document}